\documentclass[a4paper,12pt]{article}
\pdfoutput=1

\usepackage{jheppub}

\usepackage{a4wide,epsfig,psfrag,amsmath,amssymb,scalefnt}
\usepackage[dvipsnames]{xcolor}
\usepackage{amsmath,comment,braket}
\usepackage{enumitem}
\usepackage{placeins}
\usepackage{breqn}
\usepackage{slashed}
\usepackage{comment}
\usepackage{subcaption}
\usepackage{hyperref}
\usepackage{url}

\usepackage{listings}
\allowdisplaybreaks

\begin{document}

\preprint{CERN-TH-2026-051, P3H-26-019, TTP26-008}

\title{$gg\to ZH$ at NLO matched to parton showers with \texttt{ggxy} and \texttt{POWHEG}}

\author[a]{Joshua Davies,}
\author[b]{Kay Sch\"onwald,}
\author[c]{Matthias Steinhauser,}
\author[c]{Daniel Stremmer}

\emailAdd{
J.O.Davies@liverpool.ac.uk,
kay.schonwald@cern.ch,
matthias.steinhauser@kit.edu,
daniel.stremmer@kit.edu}

\affiliation[a]{Department of Mathematical Sciences, University of
    Liverpool, Liverpool, L69 3BX, UK}
\affiliation[b]{Theoretical Physics Department, CERN, 1211 Geneva 23, Switzerland}
\affiliation[c]{Institut f{\"u}r Theoretische Teilchenphysik, Karlsruhe Institute of Technology (KIT),  Wolfgang-Gaede Stra\ss{}e 1, 76131 Karlsruhe, Germany}

\abstract{

We implement the recently-calculated analytic expressions for the
next-to-leading order QCD corrections to $gg\to ZH$ in
\texttt{ggxy}. This provides a flexible framework for investigating
partonic and hadronic cross sections for various top quark mass
renormalization schemes. We augment the $Z$ boson with leptonic
decays, including spin correlations and off-shell effects, and furthermore provide an interface to \texttt{POWHEG}. This
enables simulations with parton showers, performed using
\texttt{Pythia}.

\vspace*{2em}

}

\maketitle
\flushbottom

\newpage

\vspace*{2em}

\section{Introduction}

The associated production of a Higgs boson with a gauge boson is an
important process for investigating the properties of the Higgs
boson. The experimental precision obtained at the Large Hadron Collider (LHC) is steadily
increasing
(see, e.g., Refs.~\cite{ATLAS:2024yzu,CMS:2023vzh}) which demands the continuous increase
in precision of theory predictions.

Whereas the associated $WH$ production is only possible through a
Drell-Yan type process, for $ZH$ production there is in addition a
gluon-induced channel, which is numerically important.  In recent
years many higher-order corrections have been computed both for the
total cross section and for differential
distributions~\cite{Kniehl:1990iva,Dicus:1988yh,Ciccolini:2003jy,Ferrera:2014lca,Campbell:2016jau,Hasselhuhn:2016rqt,Davies:2020drs,Wang:2021rxu,Alasfar:2021ppe,Chen:2020gae,Chen:2022rua,Davies:2025out,Davies:2025otz}
including the resummation of soft gluon effects~\cite{Kumar:2014uwa,Harlander:2014wda,Das:2025wbj}.
Most of them are implemented in the publicly available computer codes
\texttt{vh@nnlo}~\cite{Brein:2012ne,Harlander:2018yio} and
\texttt{HAWK}~\cite{Denner:2011id,Denner:2014cla}.  
In this work we
concentrate on the process $gg\to ZH$ and implement it in the
library \texttt{ggxy}~\cite{Davies:2025qjr} including next-to-leading
order (NLO) QCD corrections which are, to date, not available in public
computer codes. \texttt{ggxy} is written in \texttt{C++}, and due to its
modular structure it provides a fast and flexible framework to compute
partonic and hadronic cross sections. It is possible to choose either
the pole or the $\overline{\rm MS}$ scheme for the top quark mass. The implementation closely follows that of $gg\to HH$ at NLO QCD, which is described in
Ref.~\cite{Davies:2025qjr}.

The results discussed in this paper are based on the analytic
expressions obtained in the large top quark mass limit~\cite{Davies:2025otz}, around the
forward limit, and for high energies~\cite{Davies:2025out}.
In Section~\ref{sec::impl} we describe the implementation in
\texttt{ggxy}~\cite{Davies:2025qjr} and report on the validation against known results. We
also describe the implementation of leptonic $Z$ boson decays, including off-shell effects and spin correlations. In
Section~\ref{sec::hadr} we show how hadronic cross sections are
computed for both stable and decaying $Z$ bosons. Example files show
how \texttt{ggxy} can be used to reproduce results from the literature
available for stable $Z$ bosons.  For stable $Z$ bosons we can combine
our results with the Drell-Yan contributions from \texttt{vh@nnlo},
which is also discussed in Section~\ref{sec::hadr}.

Section~\ref{sec::powheg} is dedicated to the interface of
\texttt{ggxy} to \texttt{POWHEG}, again with the option
for stable or unstable $Z$ bosons. We validate against~\texttt{ggxy}
and then move on to simulations including parton showers
with \texttt{PYTHIA} ~\cite{Sjostrand:2007gs}.
Note that within the \texttt{POWHEG} framework the Higgs
boson is stable (though may be off-shell). Higgs boson decays can be 
performed together with the parton showers.
We conclude in Section~\ref{sec::concl}.
 
The example files shipped together with \texttt{ggxy} should allow
anyone to implement the process $gg \to ZH$, including NLO QCD
corrections, into their own analysis code.

The \texttt{ggxy} library can be downloaded from 
\url{https://gitlab.com/ggxy/ggxy-release}.
For the installation of \texttt{ggxy} we refer to Section~4 of the
original paper~\cite{Davies:2025qjr}.
Since the analytic expressions for the various expansions are
quite large, compilation time is approximately
an hour. It can be reduced by selecting only the
required processes (\verb|ggHH| or \verb|ggZH|)
in the file \verb|example-build.sh|.

\section{\label{sec::impl}Implementing $gg\to ZH$ to NLO QCD in \texttt{ggxy}}

For our calculation and implementation in {\tt ggxy} we consider the two-loop amplitudes, required for the NLO QCD corrections, that have been calculated in Ref.~\cite{Davies:2025out} using two complementary expansion methods; the forward and high-energy expansions. In the forward expansion we consider a hierarchy in which the Mandelstam variable $t$ and the external 
virtualities\footnote{Note that both the Higgs boson and $Z$ boson can be off-shell.} $q_Z^2$ and $q_H^2$ are the smallest scales of this process: $|t|,q_Z^2,q_H^2 \ll m_t^2,s$. In the high-energy expansion we consider $q_Z^2,q_H^2 \ll m_t^2 \ll s,|t|$. In particular, it was also shown in Ref.~\cite{Davies:2025out} that these two expansions methods cover the full phase space with high precision.

The amplitude for $gg\to ZH$ is expressed in terms of six independent form factors,
\begin{align} \label{eq:ffs}
    F_{12}^+(t,u),\: F_{12}^-(t,u),\: F_2^-(t,u),\: F_3^+(t,u),\: F_3^-(t,u),\: F_4(t,u)
    \,,
\end{align}
see also Eqs.~(34) and (36) of Ref.~\cite{Davies:2025out} for more details. We have implemented the analytic expressions of the two expansion methods for all form factors, and provide functions which automatically use the appropriate expansion following a similar logic as for 
$gg\to HH$, which is described in detail
in Ref.~\cite{Davies:2025qjr}. For $gg\to ZH$ there are a few differences which we describe in the following. First, we use different switching points between the two expansions for different form factors, which range from $200~{\rm GeV}<p_T<260~{\rm GeV}$ for any value of $\sqrt{s}$. In addition, we construct Pad\'{e} approximations following Ref.~\cite{Davies:2020lpf} not only for the high-energy but also for the forward expansions if $p_T>180~{\rm GeV}$. This improves the convergence of the series such that both expansions have a similar accuracy at the switching points. In order to improve the numerical stability of the high-energy expansion we perform an additional expansion in  $-t/s$ if this ratio is smaller than $0.05$. For small center-of-mass energies below $\sqrt{s}<200$~GeV we switch from the forward expansion to the large top-quark mass expansion to improve the numerical stability and evaluation time. The latter expansion has been implemented in \texttt{ggxy} up to three loops in Ref.~\cite{Davies:2025otz}. We note that at LO we use the exact expressions that have been implemented in combination with {\tt OneLOop}~\cite{vanHameren:2010cp} for the evaluation of the one-loop scalar integrals and for the double triangle contribution we use the expressions from Ref.~\cite{Davies:2020drs}.

The core of the implementation consists of the six form factors 
in Eq.~(\ref{eq:ffs}) at one- and two-loop order
from which the (partonic) Born cross section and virtual corrections used for the hadronic cross
sections are computed. In the following we present
the prototypes of the \texttt{C++} functions, which defines the input and output values in a unique way. The one- and two-loop form factors can be computed with
\begin{lstlisting}
complex<double> ggzhFF(int loops, string/int ff, double s, double t,
                       double mzs, double mhs, double mts,
                       double murs = ggzhFFmursDefault,
                       double muts = 0.0, unsigned scheme = 0,
                       complex<double> kappa_hzz = 1.0,
                       double dTriCoeff = 1.0),
\end{lstlisting}
where the input parameters are defined as,
\begin{description}[leftmargin=!,labelwidth=\widthof{\texttt{dTriCoeff:}}]
    \item[\texttt{loops:}] QCD loop order, 1 or 2
    \item[\texttt{ff:}] choice of form factor, either as a string \verb|"F12p","F12m","F2m","F3p","F3m","F4"| or as an integer between 1,...,6 following the same ordering
    \item[\texttt{s,t:}] Mandelstam variables
    \item[\texttt{mzs:}] squared Z boson mass, $m_Z^2$, can be off-shell
    \item[\texttt{mhs:}] squared Higgs boson mass, $m_H^2$, can be off-shell
    \item[\texttt{mts:}] squared top quark mass, $m_t^2$
    \item[\texttt{murs:}] squared renormalization scale $\mu_r^2$ 
    \item[\texttt{muts:}] squared renormalization scale for the $\overline{\rm MS}$ top quark mass $\mu_t^2$
    \item[\texttt{scheme:}] choice of renormalization scheme for top quark mass, 0 (OS) or 1 ($\overline{\rm MS}$). 
    \item[\texttt{kappa_hzz:}] corresponds to $\kappa_{HZZ}$
    which multiplies the $HZZ$ vertex. $\kappa_{HZZ}=1$ corresponds to the Standard Model value
    \item[\texttt{dTriCoeff:}] additional coefficient for the double-triangle contribution, can, e.g.,  be used to switch on and off the double-triangle contribution
\end{description}
The default value of the squared renormalization scale, \verb|murs|, is set to the pre-processor variable \verb|ggzhFFmursDefault| which is a placeholder for $\mu_r^2=-s$. In addition there are also low-level functions which implement the forward and high-energy expansions of the form factors. They are also accessible to the user, however we refrain from describing them in detail since they are not used directly in this paper.

The (finite) virtual corrections are implemented as 
${\cal V}_{\rm fin}$ (see Eqs.~(38) and~(39) of Ref.~\cite{Davies:2025out} for an explicit formula).
The \texttt{C++} implementation can be called with
\begin{lstlisting}
double ggzh2lVfin(double s, double t, double mzs, double mhs, double mts,
                  double GF, double murs = ggzhFFmursDefault,
                  double muts = 0.0, unsigned scheme = 0, 
                  double kappa_hzz = 1.0, double dTriCoeff = 1.0);  
\end{lstlisting}
where the input parameters are the same as before and additionally \verb|GF| stands for the Fermi  constant $G_F$. 
Note that in contrast to \texttt{ggzhFF}, here and in the following the masses $m_Z$ and $m_H$ have to be on-shell.
In addition, the pre-processor variable \verb|ggzhFFmursDefault| is, in this case, a placeholder for $\mu_r^2=s/4$. Alternatively, the function
\begin{lstlisting}
vector<double> ggzh2l(double s, double t, double mzs, double mhs,
                      double mts, double GF,
                      double murs = ggzhFFmursDefault,
                      double muts = 0.0, unsigned scheme = 0,
                      double kappa_hzz = 1.0, double dTriCoeff = 1.0);
\end{lstlisting}
can be used which returns a vector with the LO squared matrix element as the first and ${\cal V}_{\rm fin}$ as the second element. The one-loop squared matrix elements can also be computed with
\begin{lstlisting}
double ggzh1l(double s, double t, double mzs, double mhs, double mts,
              double GF, double kappa_hzz = 1.0);
\end{lstlisting}

For the computation of the hadronic cross sections
we use the generic code developed for $gg\to HH$~\cite{Davies:2025qjr}. We will provide more details in Section~\ref{sec::hadr} where several examples
are discussed in detail.

The implementation of the real corrections also closely follow our approach for $gg\to HH$.
We use the Catani-Seymour subtraction scheme~\cite{Catani:1996vz} and the $2\to 3$ matrix elements are calculated with {\tt Recola}~\cite{Actis:2012qn,Actis:2016mpe} in combination with {\tt Collier} \cite{Denner:2016kdg}. In addition, we use an alternative reduction to scalar integrals using the OPP reduction technique \cite{Ossola:2006us} implemented in {\tt CutTools}~\cite{Ossola:2007ax}, for phase-space points that are marked by {\tt Collier} to contain possibly unstable tensor integrals. In this approach the scalar integrals are calculated with {\tt OneLOop} \cite{vanHameren:2010cp}. 

\begin{figure}[t]
\centering
\includegraphics[trim= 20 590 20 20, width=0.8\textwidth]{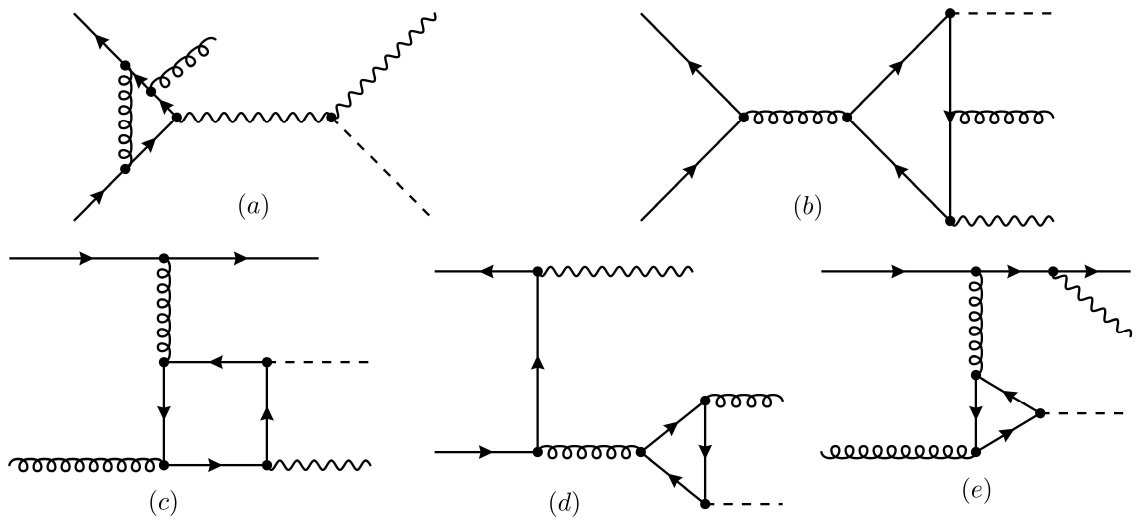}
\caption{\label{fig::sampleFDs_q_ini}
  Sample Feynman diagrams with quarks or anti-quarks in the initial state. The Drell Yan-type contributions (a) are not included in \texttt{ggxy}. 
  Using the switch \texttt{iopt_ext_Z} it is possible to include either all contributions from (b)-(e) or only those where both bosons couple to the 
  fermion loop, see (b) and (c).
  }
\end{figure}

The real corrections consist of the following sub-processes:
\begin{equation}
\begin{array}{clllcll}
gg &\to& ZHg,
&\quad&
gq/qg&\to& ZHq\,,\\
q\bar{q}/\bar{q}q &\to& ZHg,
&\quad&
g\bar{q}/\bar{q}g&\to& ZH\bar{q}\,.
\end{array}
\end{equation}
Special care has to be taken for quark-initiated sub-processes, as they contain also contributions from Higgs-Strahlung type diagrams. For the calculation of NLO QCD corrections to $gg\to ZH$ we discard all such contributions (see Fig.~\ref{fig::sampleFDs_q_ini}(a)) and take into account only contributions with a single closed quark loop.
The latter can be divided into two classes: Either
both the $Z$ and the Higgs bosons are attached to the closed quark loop, or only the Higgs boson is attached to the closed quark loop while the $Z$ is radiated from an external quark line, see Fig.~\ref{fig::sampleFDs_q_ini}(b)-(e). 
The second class of diagrams forms a gauge-invariant and finite subset and was not included in the calculation of Ref.~\cite{Chen:2022rua}, while in Refs.~\cite{Degrassi:2022mro,CampilloAveleira:2025rbh} both classes were considered. Note that the
second class has only a minor impact at the integrated level, however it can significantly impact the high-energy region of the distributions.

Since {\tt Recola} does not work with Feynman diagrams but instead generates amplitudes using the
recursive approach of Ref.~\cite{Berends:1987me}, the filtering of Higgs-Strahlung type diagrams and the separation of the two classes is not straightforward. However, we managed to eliminate the Drell-Yan-type diagrams and to implement in {\tt Recola} the possibility to obtain either only the contribution from diagrams where the $Z$ boson is attached to a closed fermion loop, or both classes of diagrams together. 
Our implementation has been cross-checked for single phase-space points with reference values from Ref.~\cite{CampilloAveleira:2025rbh}.
In \texttt{ggxy} this option is implemented with the switch \verb|iopt_ext_Z|, see the examples below for more details.

In addition to the case of stable $Z$ bosons, we have also implemented the possibility for the $Z$ boson to decay either into a pair of massless charged leptons ($Z\to\ell^-\ell^+$) or into a pair of neutrinos ($Z\to\nu_\ell\bar{\nu}_\ell$), taking into account off-shell effects and spin correlations. The corresponding amplitudes of $gg\to \ell\ell H$, can be constructed by combining the helicity amplitudes of $gg\to Z^*H$ and $Z^*\to\ell \ell$ as
\begin{align}
    \mathcal{M}(gg\to \ell\ell H)=\sum_{\lambda}\frac{\mathcal{M}_\lambda(gg\to Z^*H)\mathcal{M}_\lambda(Z^*\to\ell\ell)}{q_{Z}^2-m_Z^2+im_Z\Gamma_Z},
\end{align}
where $q_{Z}$ is the four momentum of the off-shell $Z$ boson, $m_Z$ is the on-shell $Z$ mass and $\Gamma_Z$ is the decay width of the $Z$ boson. The helicity of the $Z$ boson is indicated by $\lambda$ while the helicities of all other particles have been suppressed. The maximal value for the off-shellness of the $Z$ boson is limited in our calculation to about $150$~GeV, due to the expansion in the external virtualities of the two-loop amplitudes.
Since only the axial-vector coupling of the $Zt\bar{t}$ vertex contributes to $gg\to ZH$, while the vector component vanishes, the contribution from $\gamma^*\to\ell^-\ell^+$ originates only from the real corrections. 

For the implementation of the off-shell effects we proceed as follows: In {\tt ggxy} we first evaluate the six form factors numerically, which are used to compute the helicity amplitudes of $gg\to Z^* H$. They are then combined with the helicity amplitudes of the $Z^*$ decays. The corresponding matrix elements for the real corrections to $gg\to \ell\ell H$ can directly be obtained from {\tt Recola}. Note that here, in general, the charged lepton pair can also be produced from an off-shell photon $\gamma^*\to\ell^-\ell^+$.

We have validated the implementation
of $gg\to ZH$ in \texttt{ggxy} in various ways.
All partonic routines for the one- and two-loop
form factors and for the infra-red subtracted
NLO virtual corrections ${\cal V}_{\rm fin}$
(see Ref.~\cite{Davies:2025out} for a precise
definition)
have been cross-checked against an in-house \texttt{Mathematica} code, developed 
in parallel to Ref.~\cite{Davies:2025out}, which can evaluate using precisions
higher than \texttt{double}.

\begin{figure}[t]
\centering
\includegraphics[trim=50 20 0 20, width=0.85\textwidth]{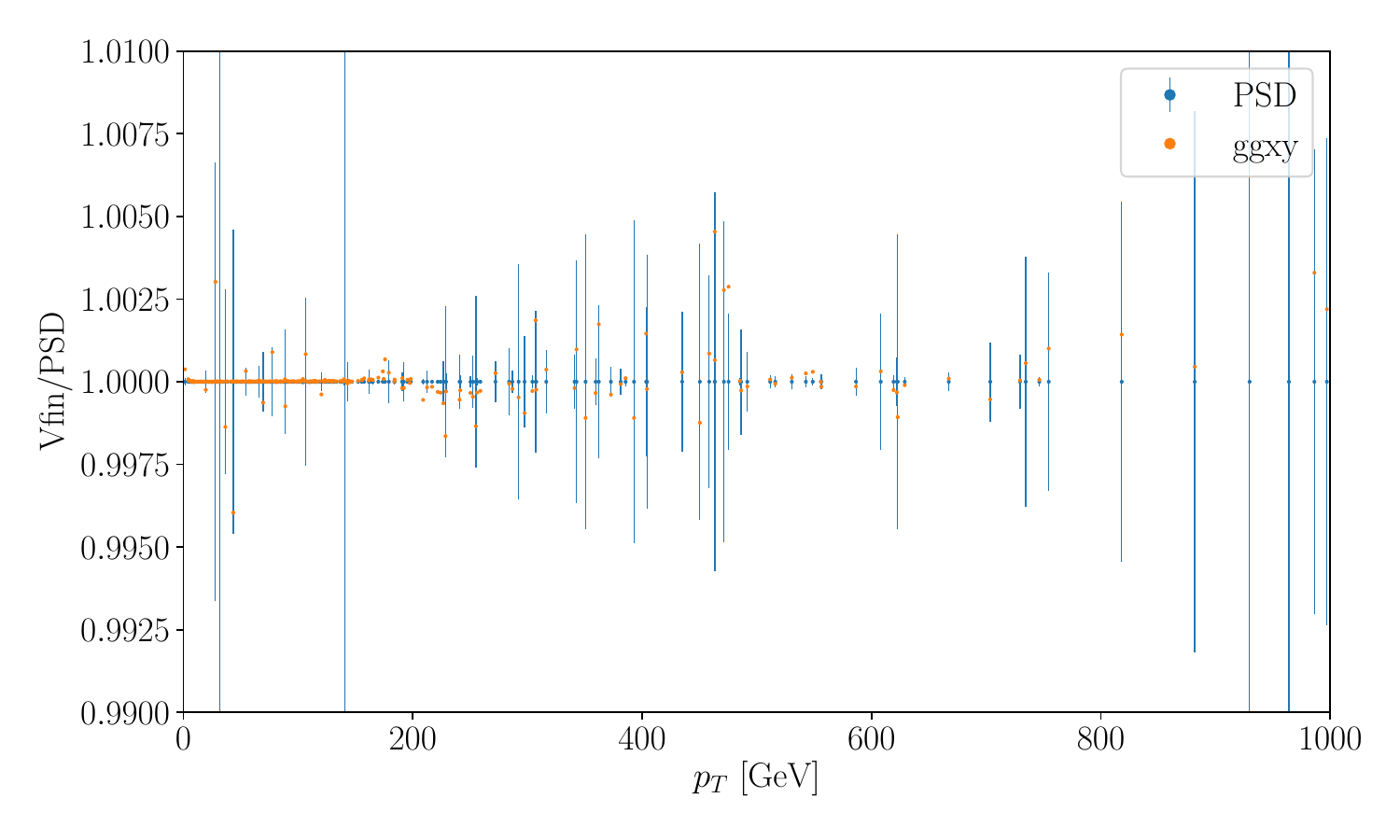}
\caption{\label{fig::vfin}
 Ratio of the finite virtual corrections
 ${\cal V}_{\rm fin}$ obtained from \texttt{ggxy} and
 Ref.~\cite{Chen:2020gae} (``PSD'') as a function
 of $p_T$. The statistical uncertainties of the reference results are shown as well.
 }
\end{figure}

In the directory \verb|examples/ggzh-FF/| we provide two examples, \verb|ff.cpp| and \verb|vfin.cpp|, which demonstrate the usage of \texttt{ggxy} for the calculation of form factors and ${\cal V}_{\rm fin}$.
After compilation using \verb|bash build.sh| one can 
call \verb|run-ff.sh| with the options ``1'' or ``2'' to obtain
results for fixed values of $\sqrt{s}$
or and $p_T$, respectively. All one- and two-loop form factors are
computed for the parameters specified in \verb|run-ff.sh|. 
The generated data files can be found in the directory \verb|dat/|
and all plots for all form factors, separated into real and imaginary parts, can be found in \verb|plots/|. 
Each plot shows the result from the forward and high-energy expansions
and the best approximation as implemented in \texttt{ggxy}.
The plots for $F_3^+$ can be compared to Figs.~14 and~15
of Ref.~\cite{Davies:2025out}. 

With the program \verb|vfin.cpp| we demonstrate the functionality and efficiency by performing a comparison of ${\cal V}_{\rm fin}$ calculated with \texttt{ggxy} with the results of Ref.~\cite{Chen:2020gae} (``PSD'') as shown in Fig.~\ref{fig::vfin}. In comparison to Fig.~16 of Ref.~\cite{Davies:2025out}, we find only minor differences which are due to a different number of expansion terms in the high-energy limit and a different switching point between the forward and high-energy expansions as described above.
As a further cross check, we have reproduced the integrated (differential) cross sections of $gg\to ZH$ at NLO QCD from Refs.~\cite{Chen:2022rua,CampilloAveleira:2025rbh}. More details will be provided in the next section.

\section{\label{sec::hadr}Hadronic fixed-order cross sections with \texttt{ggxy}}

In this section we describe how \texttt{ggxy}
can be used to compute hadronic cross
sections for $gg\to ZH$. We provide examples
where the $Z$ boson is on-shell and for the
production of off-shell $Z$ bosons which can decay into leptons. The example programs have the same structure as the one for the calculation of NLO QCD corrections to $gg\to HH$, whose implementation is described in detail in Ref.~\cite{Davies:2025qjr}.
This concerns, in particular, the information about input parameters such as the choice of the top-quark mass renormalization scheme, the PDF set, or how to define additional histograms. In the following we will
concentrate on the parts which are specific to
$gg\to ZH$.

\subsection{\label{sub::stableZ}Stable $Z$ bosons}

In the example \texttt{C++} file \verb|examples/ggzh-nlo/nlo-ggzh.cpp|
we adapt the input parameters from Ref.~\cite{CampilloAveleira:2025rbh}, which we use throughout the paper if not explicitly stated otherwise. In particular, we use $m_t=172.5~{\rm GeV}$, $m_Z=91.1876~{\rm GeV}$, $G_F=1.16637\cdot10^{-5}~{\rm GeV}^{-2}$ and the \verb|PDF4LHC21_40_pdfas| PDF set \cite{PDF4LHCWorkingGroup:2022cjn} via the LHAPDF interface \cite{Buckley:2014ana}. The renormalization and factorization scales are set to a common scale $\mu_r=\mu_f=M_{ZH}/2$, where $M_{ZH}$ is the invariant mass of the $ZH$ system. $Z$ radiation from external quark lines is enabled, and may be controlled in the example program with the variable \verb|iopt_ext_Z = 1|
or \verb|iopt_ext_Z = 0|.
After compilation (e.g.~by executing \verb|bash build.sh|) one can run
the example with the help of the 
script \verb|run.sh| which can be 
found in the same directory.
\verb|bash run.sh 1 1| generates the
histogram files in \verb|results-ZH/hist/|.
They can be processed with
\verb|bash run.sh 2 1| which produces
the LO and NLO total cross sections,
including the Monte Carlo uncertainty and the uncertainty from
the 7-point scale variation at $13.6$~TeV.
The results are shown in Tab.~\ref{tab::sig2}
together with the results from Ref.~\cite{CampilloAveleira:2025rbh}.
For convenience we also show results
for $\sqrt{s}=13$~TeV and $14$~TeV.
The comparison of the total cross section to Ref.~\cite{Chen:2022rua} is shown in Tab.~\ref{tab::sig1}, where we adapt the corresponding input values. In both cases \texttt{ggxy} is able to reproduce the reference results precisely. The example script demonstrates the cross section computation with different seeds in parallel, where the run-time for each seed is about $90$ minutes on a single core, which is three times longer than that of $gg\to HH$. The NLO values in both tables are based on the default setting of five seeds (cores) in parallel.

\begin{table}[t]
    \centering
    \renewcommand{\arraystretch}{1.2}
    \begin{tabular}{cc@{\hskip 10mm}l@{\hskip 10mm}l@{\hskip 10mm}}
        \hline
        $\sqrt{s}$&
        &{ \tt ggxy}&Ref.~\cite{CampilloAveleira:2025rbh}  \\
        \hline
        $13$~TeV & $\sigma^{\rm LO}$ [fb]& $63.97(2)^{+26.7\%}_{-20.0\%}$ &    $64.0^{+27\%}_{-20\%}$\\
        & $\sigma^{\rm NLO}$ [fb]& $118.40(8)^{+16.5\%}_{-14.0\%}$ & $118.4^{+17\%}_{-14\%}$\\
        \noalign{\smallskip}\hline\noalign{\smallskip}
        $13.6$~TeV & $\sigma^{\rm LO}$ [fb]& $70.59(2)^{+26.3\%}_{-19.7\%}$ &    $70.6^{+26\%}_{-20\%}$\\
        & $\sigma^{\rm NLO}$ [fb]& $130.50(9)^{+16.3\%}_{-13.8\%}$ & $130.5^{+16\%}_{-14\%}$\\

        \noalign{\smallskip}\hline\noalign{\smallskip}
        $14$~TeV & $\sigma^{\rm LO}$ [fb]& $75.15(2)^{+26.0\%}_{-19.6\%}$ &    $75.2^{+26\%}_{-20\%}$\\
        & $\sigma^{\rm NLO}$ [fb]& $139.0(1)^{+16.3\%}_{-13.7\%}$ & $138.9^{+16\%}_{-14\%}$\\
        \noalign{\smallskip}\hline\noalign{\smallskip}
    \end{tabular}
    \caption{\label{tab::sig2} Comparison with results of Ref.~\cite{CampilloAveleira:2025rbh}
    at $\sqrt{s}=\{13,13.6,14\}$~TeV. }
\end{table}

\begin{table}[t]
    \centering
    \renewcommand{\arraystretch}{1.2}
    \begin{tabular}{cc@{\hskip 10mm}l@{\hskip 10mm}l@{\hskip 10mm}}
        \hline
        $\sqrt{s}$&
        &{ \tt ggxy}&Ref.~\cite{Chen:2022rua}  \\
        \hline
        $13$~TeV & $\sigma^{\rm LO}$ [fb]& $52.41(2)^{+25.5\%}_{-19.3\%}$ &    $52.42^{+25.5\%}_{-19.3\%}$\\
        & $\sigma^{\rm NLO}$ [fb]& $104.23(6)^{+16.5\%}_{-13.9\%}$ & $103.8(3)^{+16.4\%}_{-13.9\%}$\\
        \noalign{\smallskip}\hline\noalign{\smallskip}
        $13.6$~TeV & $\sigma^{\rm LO}$ [fb]& $58.06(2)^{+25.1\%}_{-19.0\%}$ &    $58.06^{+25.1\%}_{-19.0\%}$\\
        & $\sigma^{\rm NLO}$ [fb]& $115.16(7)^{+16.2\%}_{-13.8\%}$ & $114.7(3)^{+16.2\%}_{-13.7\%}$\\
        \noalign{\smallskip}\hline\noalign{\smallskip}
        $14$~TeV & $\sigma^{\rm LO}$ [fb]& $61.95(2)^{+24.9\%}_{-18.9\%}$ &    $61.96^{+24.9\%}_{-18.9\%}$\\
        & $\sigma^{\rm NLO}$ [fb]& $122.64(8)^{+16.1\%}_{-13.7\%}$ & $122.2(3)^{+16.1\%}_{-13.6\%}$\\        
        \noalign{\smallskip}\hline\noalign{\smallskip}
    \end{tabular}
    \caption{\label{tab::sig1} Comparison with results of Ref.~\cite{Chen:2022rua}
    at $\sqrt{s}=\{13,13.6,14\}$~TeV.}
\end{table}

\begin{figure}[t]
  \begin{center}
  \begin{tabular}{cc}
   \includegraphics[width=.45\textwidth]{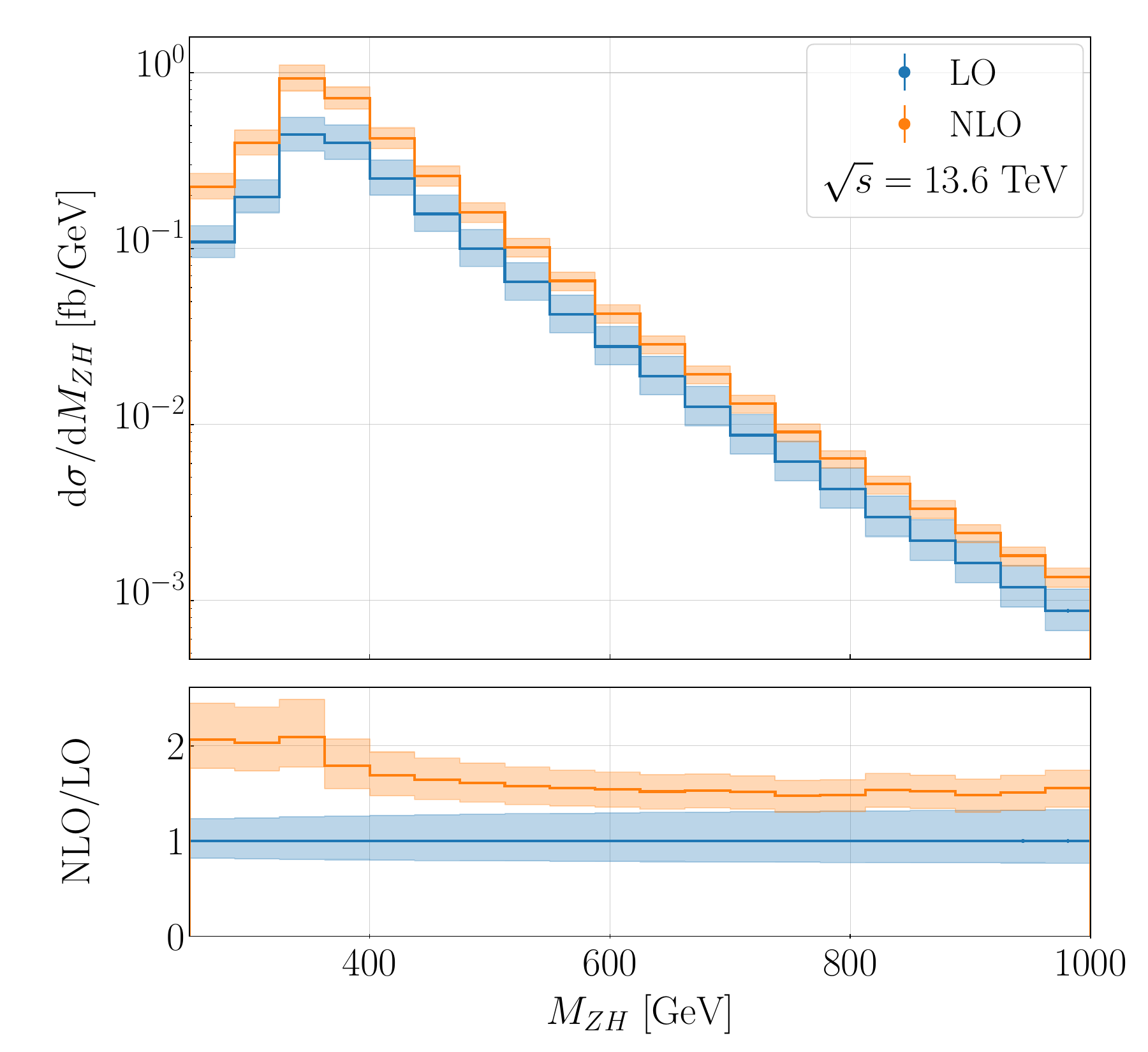} &
   \includegraphics[width=.45\textwidth]{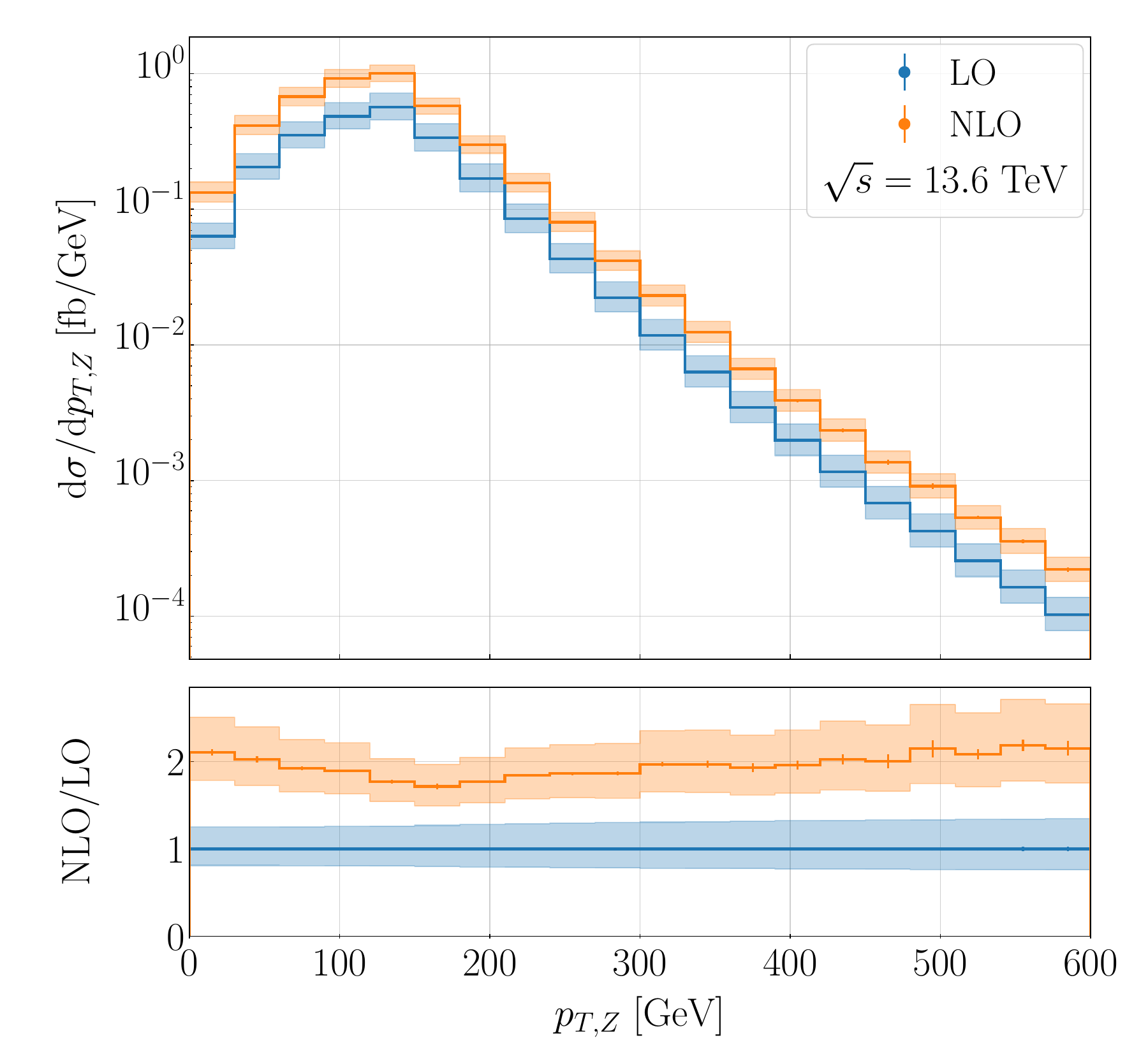}
  \end{tabular}
  \end{center}
  \caption{\label{fig::kfactor_ggxy_ZH}
  $M_{ZH}$ invariant mass and $Z$ boson transverse momentum distribution for stable $Z$ bosons. The input parameters
  from Ref.~\cite{CampilloAveleira:2025rbh} have been adapted.}
\end{figure}

Running \verb|bash run.sh 2 1| also generates plots for several distributions, which can be found in the directory \verb|results/hist/kfactor/|.
In Fig.~\ref{fig::kfactor_ggxy_ZH} we show, as examples, the $ZH$ invariant mass distribution and the transverse momentum distribution of the $Z$ boson. 
It is straightforward to adapt the file \verb|examples/ggzh-nlo/nlo-ggzh.cpp| to one's own requirements, see also Ref.~\cite{Davies:2025qjr}.

\begin{figure}[t]
  \begin{center}
  \begin{tabular}{cc}
   \includegraphics[width=.45\textwidth]{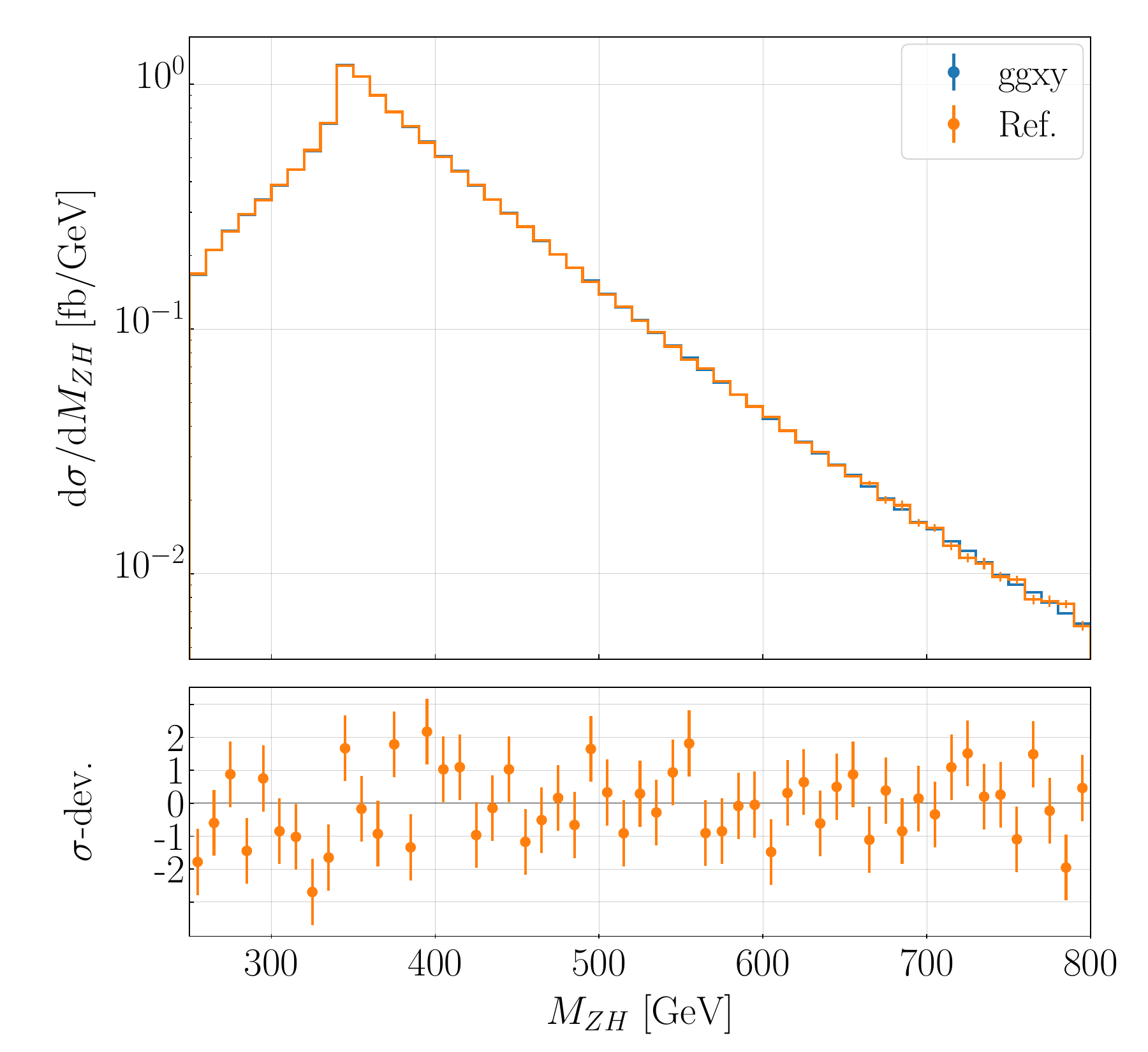} &
   \includegraphics[width=.45\textwidth]{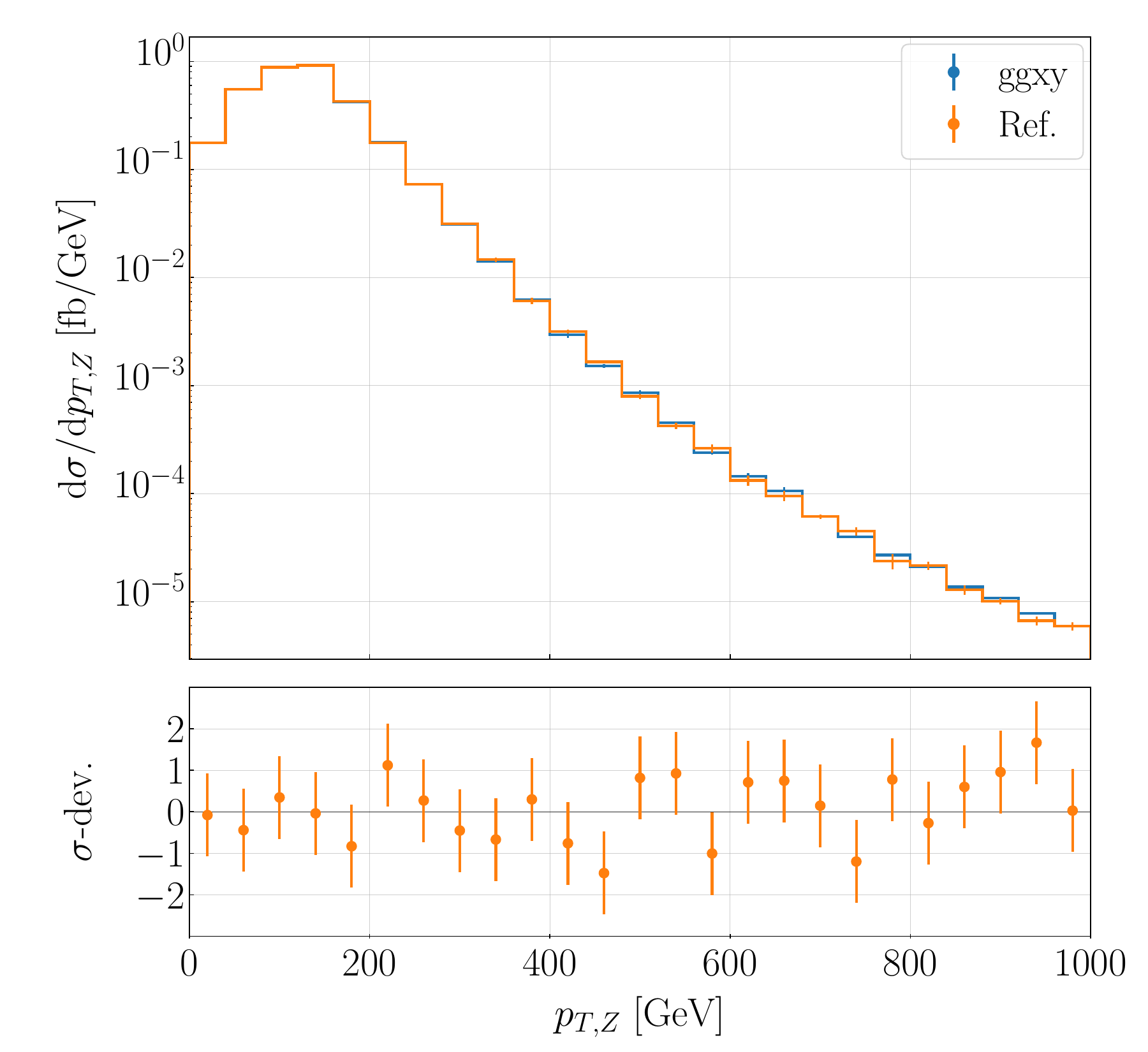}
  \end{tabular}
  \end{center}
  \caption{\label{fig::comparison_ggxy_ZH}
  Comparison with Ref. \cite{CampilloAveleira:2025rbh} for the distributions in $M_{ZH}$ and $p_{T,Z}$.
  In the lower panels we show the difference of the results normalized to the combined statistical uncertainties.}
\end{figure}

In Fig.~\ref{fig::comparison_ggxy_ZH}, we show the comparison with the results from Ref.~\cite{CampilloAveleira:2025rbh} for the distributions in $M_{ZH}$ and $p_{T,Z}$. The lower panels present the standard deviations with respect to the statistical uncertainties,
which shows good agreement between both predictions.

\begin{figure}[t]
  \begin{center}
  \begin{tabular}{cc}
   \includegraphics[width=.45\textwidth]{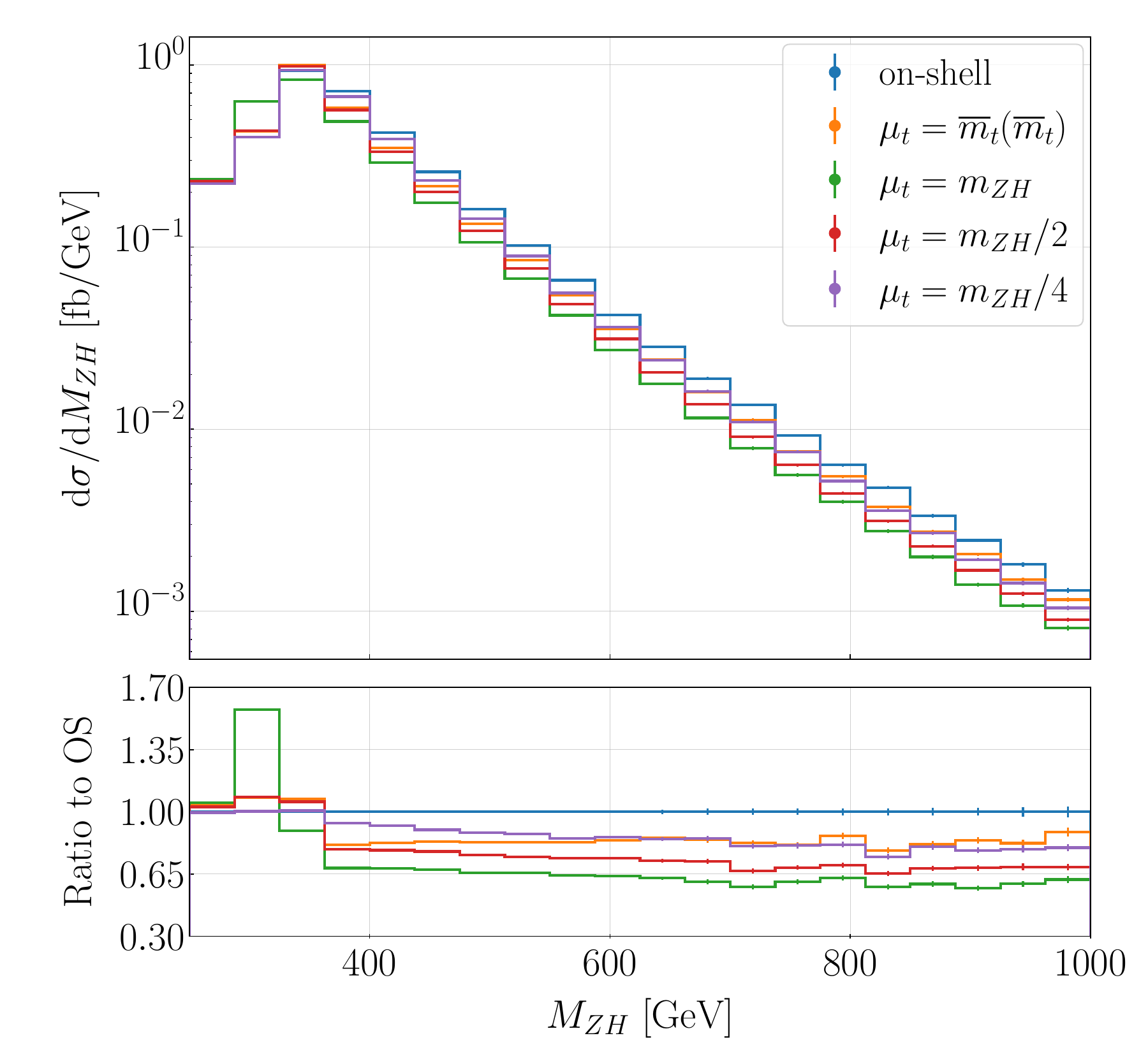} &
   \includegraphics[width=.45\textwidth]{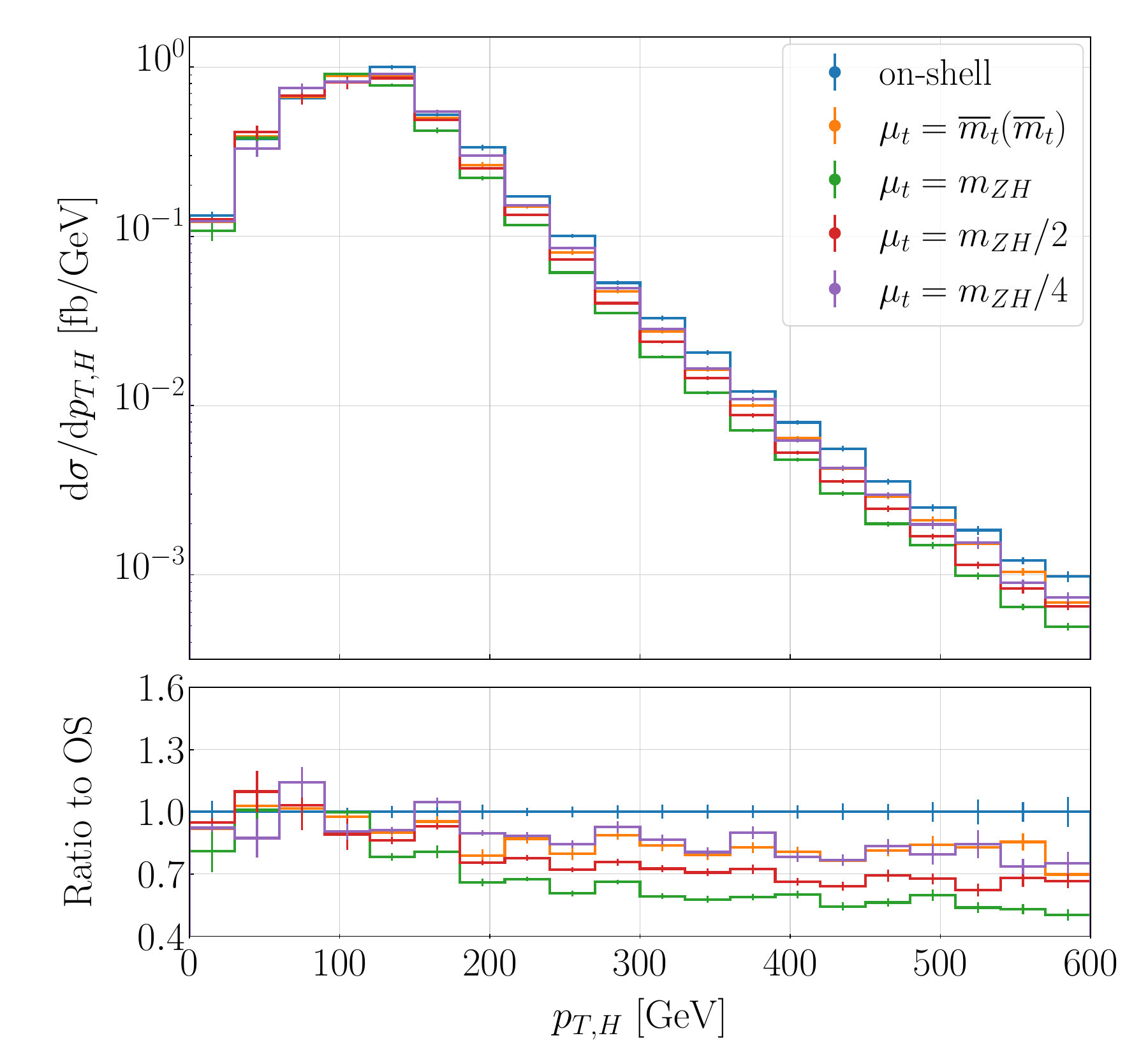}
  \end{tabular}
  \end{center}
  \caption{\label{fig::OSvsMS}
  Dependence on the top quark mass renormalization scale for the distributions in $M_{ZH}$ and $p_{T,H}$.
  In the lower panels we show the ratios between the different choices of $\overline{\text{MS}}$ scales to the on-shell scheme for fixed $\mu_f=\mu_r=M_{ZH}/2$.}
\end{figure}

It is possible to change the renormalization between the OS and $\overline{\text{MS}}$ schemes by setting \verb|scheme| to 0 or 1, respectively.
In the latter case one has to provide the renormalization scale of the top quark mass $\mu_t^2$, which can either be a fixed or a dynamical scale.
In Fig.~\ref{fig::OSvsMS} we show the dependence of the distributions in $M_{ZH}$ and $p_{T,H}$ on the choice of top quark mass renormalization scale.
We typically observe a downward shift for large invariant mass of the order of $10-40\,$\%, in line with what has been observed in earlier work on $gg \to ZH$~\cite{Degrassi:2022mro}.
The variation of the top mass renormalization scheme therefore introduces an uncertainty of a similar size to that of the scale variation of $\mu_r$ and $\mu_f$.
In the following we choose to work with the on-shell mass.

\subsection{\label{sub::ggxy_offZ}Off-shell $Z$ bosons and leptonic decays}

An example for the computation of NLO QCD corrections to off-shell $Z$ production with leptonic decays can be found in the file \verb|examples/ggzh-nlo/nlo-ggllh.cpp|, where again we adopt the input parameters from
Ref.~\cite{CampilloAveleira:2025rbh}. The final state is controlled by the variable \verb|decaymode| with the values \verb|"ll"| for $gg\to\ell^-\ell^+H$ (default) and \verb|"vv"| for $gg\to\nu_\ell\bar{\nu}_\ell H$.\footnote{All results calculated with the example program are for a single massless lepton generation, and can be multiplied by $2$ to consider also $\ell=e,\mu$ or $3$ for $\ell=e,\mu,\tau$.} In addition, the two parameters \verb|mllmin| and \verb|mllmax| control the size of the invariant mass of the two-lepton system as $\verb|mllmin|<M_{\ell\ell}<\verb|mllmax|$. A lower limit is only necessary for $gg\to\ell^-\ell^+H$ since in this case the two charged leptons can be produced from an off-shell photon in the real corrections. We use \verb|mllmin = 15.0 (GeV)| as the default. The default value of \verb|mllmax| is set to $150$~GeV and should not be increased due to the expansions in the external virtualities of the virtual amplitudes. Compilation and execution of the sample file proceed 
as in Section~\ref{sub::stableZ} where the second input parameter of
\verb|run.sh| is ``2'' instead of ``1''. The inclusion of $Z$ decays has no impact on the runtime.

\begin{table}[t]
    \centering
    \renewcommand{\arraystretch}{1.2}
    \begin{tabular}{cc@{\hskip 10mm}l@{\hskip 10mm}l@{\hskip 10mm}}
        \hline
        $\sqrt{s}$&
        &$gg\to\ell^-\ell^+H$&$gg\to\nu_\ell\bar{\nu}_\ell H$  \\
        \hline
        $13$~TeV & $\sigma^{\rm LO}$ [fb]& $6.366(4)^{+26.7\%}_{-20.0\%}$ & $12.581(8)^{+26.7\%}_{-20.0\%}$\\
        & $\sigma^{\rm NLO}$ [fb]& $11.80(1)^{+16.6\%}_{-14.0\%}$ & $23.30(2)^{+16.5\%}_{-14.0\%}$\\
        \noalign{\smallskip}\hline\noalign{\smallskip}
        $13.6$~TeV & $\sigma^{\rm LO}$ [fb]& $7.026(4)^{+26.2\%}_{-19.7\%}$ & $13.883(8)^{+26.2\%}_{-19.7\%}$\\
        & $\sigma^{\rm NLO}$ [fb]& $13.02(1)^{+16.4\%}_{-13.8\%}$ & $25.70(2)^{+16.4\%}_{-13.8\%}$\\

        \noalign{\smallskip}\hline\noalign{\smallskip}
        $14$~TeV & $\sigma^{\rm LO}$ [fb]& $7.477(5)^{+26.0\%}_{-19.6\%}$ & $14.779(9)^{+26.0\%}_{-19.6\%}$\\
        & $\sigma^{\rm NLO}$ [fb]& $13.84(1)^{+16.3\%}_{-13.8\%}$ & $27.30(2)^{+16.3\%}_{-13.7\%}$\\
        \noalign{\smallskip}\hline\noalign{\smallskip}
    \end{tabular}
    \caption{\label{tab::sig3} Results for off-shell $Z$ production with leptonic $Z$ decays
    at $\sqrt{s}=\{13,13.6,14\}$~TeV.}
\end{table}

In Tab.~\ref{tab::sig3} we present results for the integrated cross section for $gg\to\ell^-\ell^+H$ and $gg\to\nu_\ell\bar{\nu}_\ell H$ at $\sqrt{s}=\{13,13.6,14\}$~TeV, where in both cases we consider $\ell=e,\mu,\tau$. The inclusion of (leptonic) $Z$ decays has no impact on the size of NLO QCD corrections at the integrated level which are of the order of $85\%$, as can be seen from the comparison with Tab.~\ref{tab::sig2}.

\begin{figure}[t]
  \begin{tabular}{cc}
   \includegraphics[width=.45\textwidth]{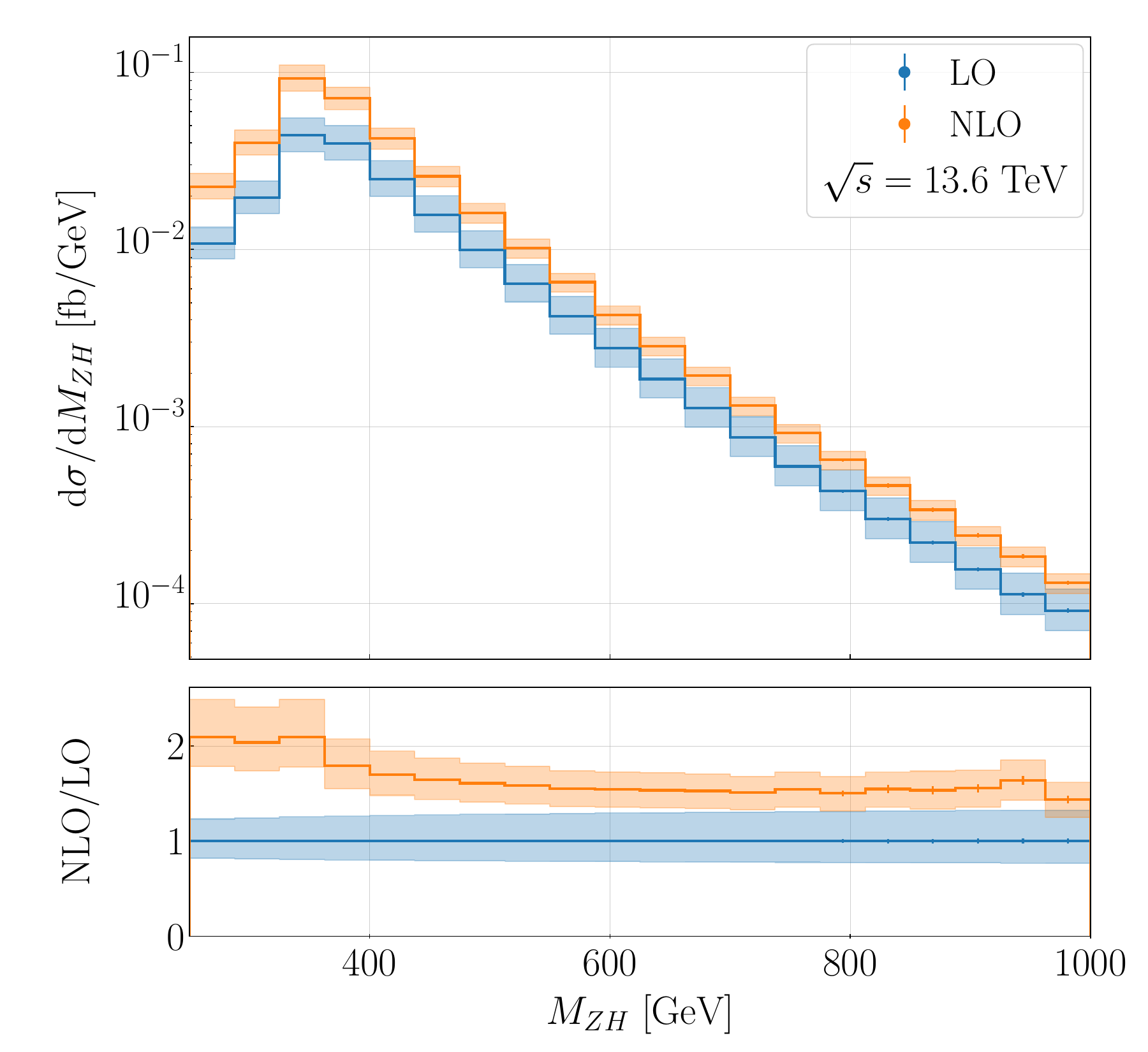} &
   \includegraphics[width=.45\textwidth]{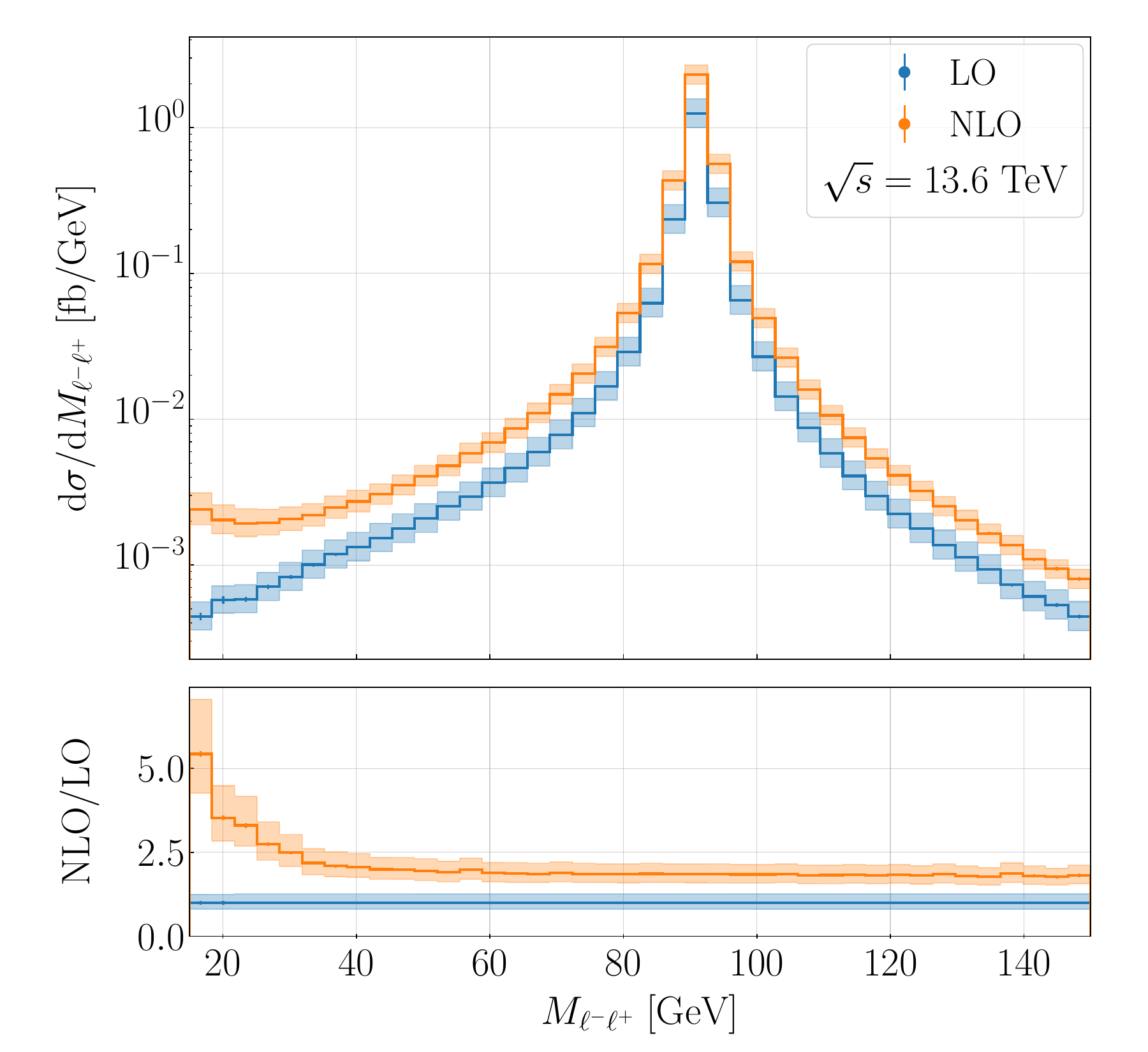}
  \end{tabular}
  \caption{\label{fig::compare_ggxy_llH}
  LO and NLO QCD predictions for $M_{ZH}$ and $M_{\ell^-\ell^+}$ for the $gg\to\ell^-\ell^+H$ process.}
\end{figure}

In Fig.~\ref{fig::compare_ggxy_llH} we show as
an example for the process $gg\to\ell^-\ell^+H$ the $ZH$ invariant mass and the lepton pair invariant mass. While the former observable is essentially identical to the counterpart shown before for a stable $Z$ boson, in the latter case the $Z$ boson peak is clearly visible as well as the enhancement towards small invariant masses $M_{\ell^-\ell^+}$ due to $\gamma^\star\to\ell^-\ell^+$ starting at NLO. 
Altogether, after running the example as described above, plots for 16 different distributions are generated.
They can be found in the directory
\verb|examples/ggzh-nlo/results-llH/kfactor|.

\subsection{Combination with \texttt{vh@nnlo}}

Our results for the gluon fusion process can be combined with the
Drell-Yan process for $ZH$.  It starts at LO and is known to
N$^3$LO~\cite{Baglio:2022wzu}; corrections up to NNLO are implemented in the
program \verb|vh@nnlo|~\cite{Brein:2012ne,Harlander:2018yio} based on work presented in~\cite{Han:1991ia,Hamberg:1990np,Harlander:2002wh,Brein:2003wg,Ciccolini:2003jy,Brein:2011vx,Kniehl:1990iva,Altenkamp:2012sx,Harlander:2013mla}.
In the following we use \verb|vh@nnlo| to compute the
total cross section.

\begin{table}[t]
    \centering
    \renewcommand{\arraystretch}{1.2}
    \begin{tabular}{cc@{\hskip 10mm}l@{\hskip 10mm}}
        \hline
        $\sqrt{s}$ & &$\rm vh@nnlo$  \\
        \hline
        $13$~TeV &
        $\sigma_{\rm top}$[fb] & $122.7(7)^{+15.6\%}_{-13.6\%}$
        \\
        & $\sigma_{\rm DY}$ [fb]& $801.9(2)^{+0.14\%}_{-0.25\%}$ 
        \\
        & $\sigma_{\rm I + II}$ [fb]& $11.30(7)^{+21.9\%}_{-16.9\%}$
        \\
        & $\sigma_{b \bar{b} \to ZH}$ [fb]& $0.3793(6)^{+16.3\%}_{-18.5\%}$
        \\
        & $\sigma_{\rm vh@nnlo}$ [fb]& $813.6(2)^{+0.32\%}_{-0.49\%}$ 
        \\
        \hline
        $13.6$~TeV &
        $\sigma_{\rm top}$[fb] & $135.3(8)^{+15.7\%}_{-13.4\%}$
        \\
        & $\sigma_{\rm DY}$ [fb]& $852.3(2)^{+0.14\%}_{-0.25\%}$ 
        \\
        & $\sigma_{\rm I + II}$ [fb]& $12.01(8)^{+23.4\%}_{-16.4\%}$
        \\
        & $\sigma_{b \bar{b} \to ZH}$ [fb]& $0.4218(7)^{+16.5\%}_{-18.6\%}$
        \\
        & $\sigma_{\rm vh@nnlo}$ [fb]& $864.7(2)^{+0.34\%}_{-0.48\%}$ 
        \\
        \hline
        $14$~TeV &
        $\sigma_{\rm top}$[fb] & $144.0(8)^{+15.7\%}_{-13.4\%}$
        \\
        & $\sigma_{\rm DY}$ [fb]& $886.2(2)^{+0.14\%}_{-0.25\%}$ 
        \\
        & $\sigma_{\rm I + II}$ [fb]& $12.62(9)^{+22.9\%}_{-17.1\%}$
        \\
        & $\sigma_{b \bar{b} \to ZH}$ [fb]& $0.4513(7)^{+16.6\%}_{-18.7\%}$
        \\
        & $\sigma_{\rm vh@nnlo}$ [fb]& $899.2(3)^{+0.34\%}_{-0.49\%}$ 
        \\ \noalign{\smallskip}\hline\noalign{\smallskip}
    \end{tabular}
    \caption{\label{tab::vh@nnlo} Results for vh@nnlo at $\sqrt{s}=\{13,13.6,14\}$~TeV.
    $\sigma_{\rm top}$ is the reweighted $gg \to ZH$ contributions, which we replace with our calculation.
    The Drell-Yan like contributions are given by $\sigma_{\rm DY}$.
    Top quark effects in the Drell-Yan like contributions are given by $\sigma_{\rm I + II}$.
    Bottom quark induced processes are given by $\sigma_{b \bar{b} \to ZH}$.
    We define $\sigma_{\rm vh@nnlo}=\sigma_{\rm DY}+\sigma_{\rm I + II}+\sigma_{b \bar{b} \to ZH}$ as the contributions to which we want to add to our new predictions, i.e.~omitting $\sigma_{\rm top}$.
    }
\end{table}

In Tab.~\ref{tab::vh@nnlo} we show the output 
of \verb|vh@nnlo| for $\sqrt{s}=13,13.6$ and $14$~TeV
separated into the individual contributions
$\sigma_{\rm top}$,
$\sigma_{\rm DY}$,
$\sigma_{\rm I + II}$,
$\sigma_{b \bar{b} \to ZH}$,
and $\sigma_{\rm vh@nnlo}$.
The first cross section corresponds to the production channel $gg \to ZH$, where the exact leading order is rescaled with the $K$-factor obtained from the infinite top mass limit at NLO,
the second represents the Drell-Yan like topologies without heavy quark loops,
the third to $q \bar{q}$ induced contributions involving the heavy top quark, 
and the forth to $b \bar{b}$ induced contributions, where the bottom quark is treated as massless, but the Yukawa coupling to the Higgs boson is retained.
$\sigma_{\rm vh@nnlo} = \sigma_{\rm DY}+\sigma_{\rm I + II}+\sigma_{b \bar{b} \to ZH}$ is the sum of all contributions which are not implemented in \texttt{ggxy} and therefore can be safely combined with our implementation of $gg \to ZH$.

\begin{figure}[t]
\centering
\includegraphics[width=.85\textwidth]{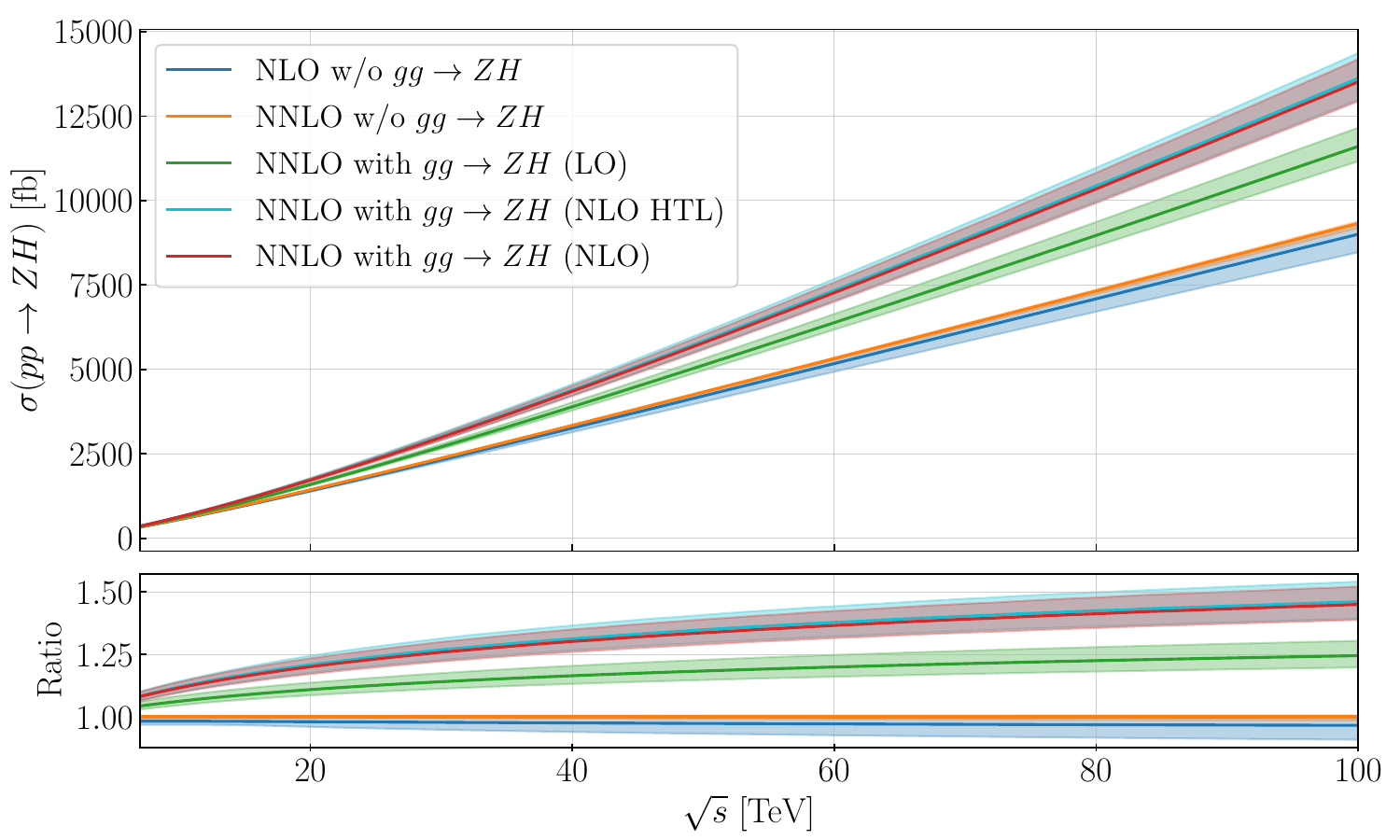}
\caption{\label{fig::cross_ppZH}Total cross section for $ZH$ production as a function of the collider energy up to $\sqrt{s}=100$~TeV. We separately show the cross section at NLO and NNLO without the gluon-induced channel and at NNLO including LO and also NLO contributions from $gg\to ZH$.}
\end{figure}

In Fig.~\ref{fig::cross_ppZH} we show the total $ZH$ production cross
section as a function of the collider energy between 7 and 100~TeV.
The dark blue and orange curves correspond to the Drell-Yan contribution at NLO and NNLO, including the uncertainty from the variation of the renormalization
and factorization scales, which increases (at NNLO) from $0.5\%$ to $1.5\%$ for higher values of $\sqrt{s}$. This is in line with~\cite{Baglio:2022wzu} (see also
Tab.~\ref{tab::vh@nnlo}). This contribution does not include the
gluon-induced processes, however, it includes
the contributions $\sigma_{I+II}$ and $\sigma_{b \bar{b} \to ZH}$
(see above).
For simplicity, we do not       
consider electroweak corrections since they are 
implemented in \texttt{vh@nnlo} only up to $\sqrt{s}=14$~TeV.

The green and red bands in Fig.~\ref{fig::cross_ppZH} show the
total cross section including the gluon-induced channel,  where the
green band only includes LO result and the red also the NLO
corrections. The size of the gluon-fusion contribution increases from $5\%$ to $25\%$ of the DY contribution for rising values of $\sqrt{s}$. The NLO cross section is about twice the size. At LHC energies we observe a $15\%$ effect of the $gg$ channel. By including the NLO corrections, the scale uncertainties from the gluon-fusion contribution decrease from $22\%-33\%$ to $19\%-13\%$, where the scale uncertainties decreases towards higher values of $\sqrt{s}$. The inclusion of the $gg$ channel in $ZH$ production increases the combined scale uncertainties to $2\%-5\%$.

For the $gg\to ZH$ channel, we also show the NLO corrections as obtained from the rescaling of the 
exact LO result with the $K$ factor obtained from the NLO
heavy-top contribution, in the light-blue curve.
We observe that the deviation from the exact NLO result
is between $1\%-2\%$ over the whole energy range.
Differential distributions usually show larger deviations from the naive heavy top quark limit, see for example the discussion for the similar $gg\to HH$ process in Ref.~\cite{Borowka:2016ypz}.
The NLO corrections are large and need to be taken into account.

In Ref.~\cite{Baglio:2022wzu} the N3LO DY contributions have been calculated, and amount to about $1\%-2\%$, while the scale uncertainties barely change.
Since for the total cross section the scale uncertainties are dominated by the $gg\to ZH$ process and the rescaling by the heavy top quark mass limit delivers reliable results at NLO, this motivates the full calculation of NNLO corrections to $gg\to ZH$ in this limit. The virtual corrections are already known and are given in Ref.~\cite{Davies:2025otz}.


\section{\label{sec::powheg} $gg\to ZH$ at NLO matched to a parton shower}

\subsection{Implementation in \texttt{POWHEG}: \texttt{ggxy_ggZH}}

In the following we describe the new implementation of $gg\to ZH$, including NLO QCD corrections,
in \texttt{POWHEG}~\cite{Alioli:2010xd} via an interface to \texttt{ggxy}, which allows the combination with parton showers from e.g.~\texttt{Pythia}~\cite{Sjostrand:2007gs} or \texttt{Herwig}~\cite{Bellm:2025pcw}. We note that this interface is based on the \texttt{POWHEG} process implementation of $gg\to HH.$\footnote{\url{https://gitlab.com/POWHEG-BOX/V2/User-Processes/ggxy_ggHH}} For $gg\to ZH$ we consider a leptonically-decaying $Z$ boson, into either a charged lepton or neutrino pair, taking into account spin correlations as well as off-shell effects at the matrix element level. This code can be obtained from \url{https://gitlab.com/POWHEG-BOX/V2/User-Processes/ggxy_ggZH}. 

In an alternative approach (primarily for testing purposes) we have also implemented $gg\to ZH$ with a stable $Z$ boson.\footnote{This implementation can be download from \url{https://gitlab.com/ggxy/ggxy_ggZH-stable}.} In this case, the $Z$ decays can be generated with the parton shower Monte Carlo program. Differences of both implementations at the differential level will be discussed in Section~\ref{sub::num}.

All one-loop matrix elements are calculated with \texttt{Recola}, while the virtual corrections are obtained from \texttt{ggxy}. It is thus necessary first to install \texttt{ggxy} and to define the path to the directory \verb|example-install| of \texttt{ggxy} in the variable \verb|path_to_ggxy| in the file \verb|env-vars.sh|,
which can be found in the \texttt{POWHEG} directory. The \texttt{POWHEG} process can then be compiled with \verb|source env-vars.sh| followed by \verb|make|, for more details see also the \verb|README.md| of the process implementation.

This implementation is as flexible as the one which only uses \texttt{ggxy}, see the previous sections, and allows for the variation of all input parameters as well as the top-quark mass renormalization scheme. The invariant mass range of the two lepton/neutrino system can be specified with the variables \verb|min_z_mass| and \verb|max_z_mass|, where the default values are set to $15~{\rm GeV}$ and $150~{\rm GeV}$, respectively. Smaller values of \verb|min_z_mass| might lead to a less efficient event generation, and the value of \verb|max_z_mass| should not be increased as the two-loop amplitudes are expansions in the virtualities of the (off-shell) $Z$ and Higgs bosons.  To improve the numerical stability, we have implemented a technical cut where we discard events in the real corrections that contain a radiated parton with a transverse momentum of less than $0.1$~GeV, see also Ref.~\cite{Alioli:2021wpn}.
Further information about process-dependent parameters of the POWHEG implementation can be found in the \texttt{README.md} file.

After installation, an example setup can be found in the directory \verb|testrun|, which can be started with \verb|run.sh|. Input parameters should be adjusted beforehand in the file \verb|powheg.input-save|. The script \verb|run.sh| runs through all four stages of \texttt{POWHEG}, where the first stage
is used to optimize the phase-space generation. Differential distributions and the total cross section are calculated in the second stage. The output of stage two represents the fixed-order results and can be directly compared with the output of \texttt{ggxy}. Stage~3 is used to initialize the event generation that is carried out in stage~4.

\subsection{Cross checks with ggxy}

We have performed several cross checks of the fixed-order results from \texttt{POWHEG} after the end of stage~2 with the results from \texttt{ggxy}. For stable $Z$ bosons we reproduce the integrated cross section at $\sqrt{s}=13.6~{\rm TeV}$ from Tab.~\ref{tab::sig2} and obtain
\begin{align*}
    \sigma^{\rm NLO}_{gg\to ZH}(\texttt{ggxy})&= 130.50(9)~{\rm fb},\\
    \sigma^{\rm NLO}_{gg\to ZH}(\texttt{POWHEG})&= 130.56(4)~{\rm fb},
\end{align*}
where we find an agreement between the two calculation within less than one standard deviation.
\begin{figure}[t]
\centering
\includegraphics[width=.32\textwidth]{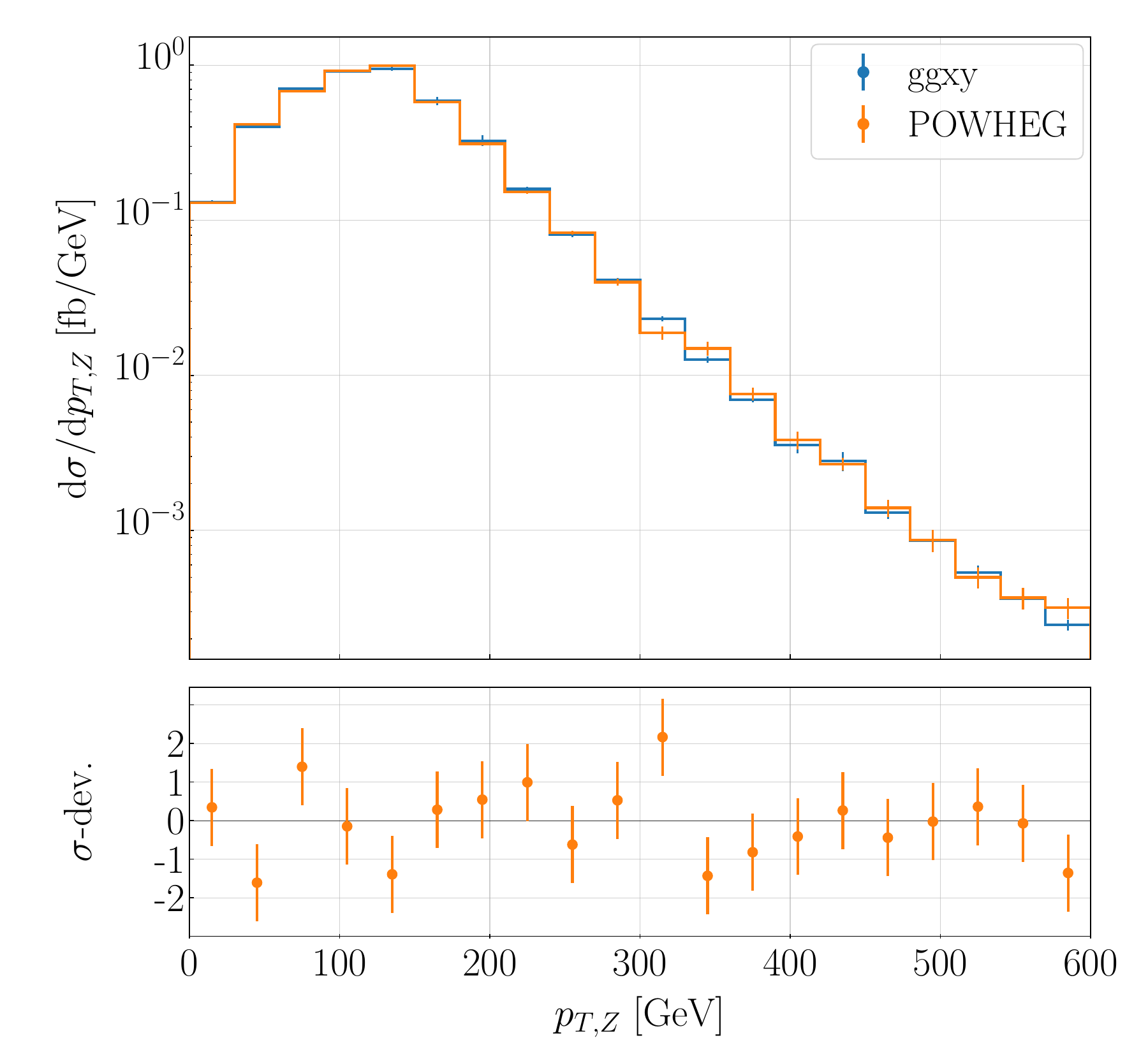}
\includegraphics[width=.32\textwidth]{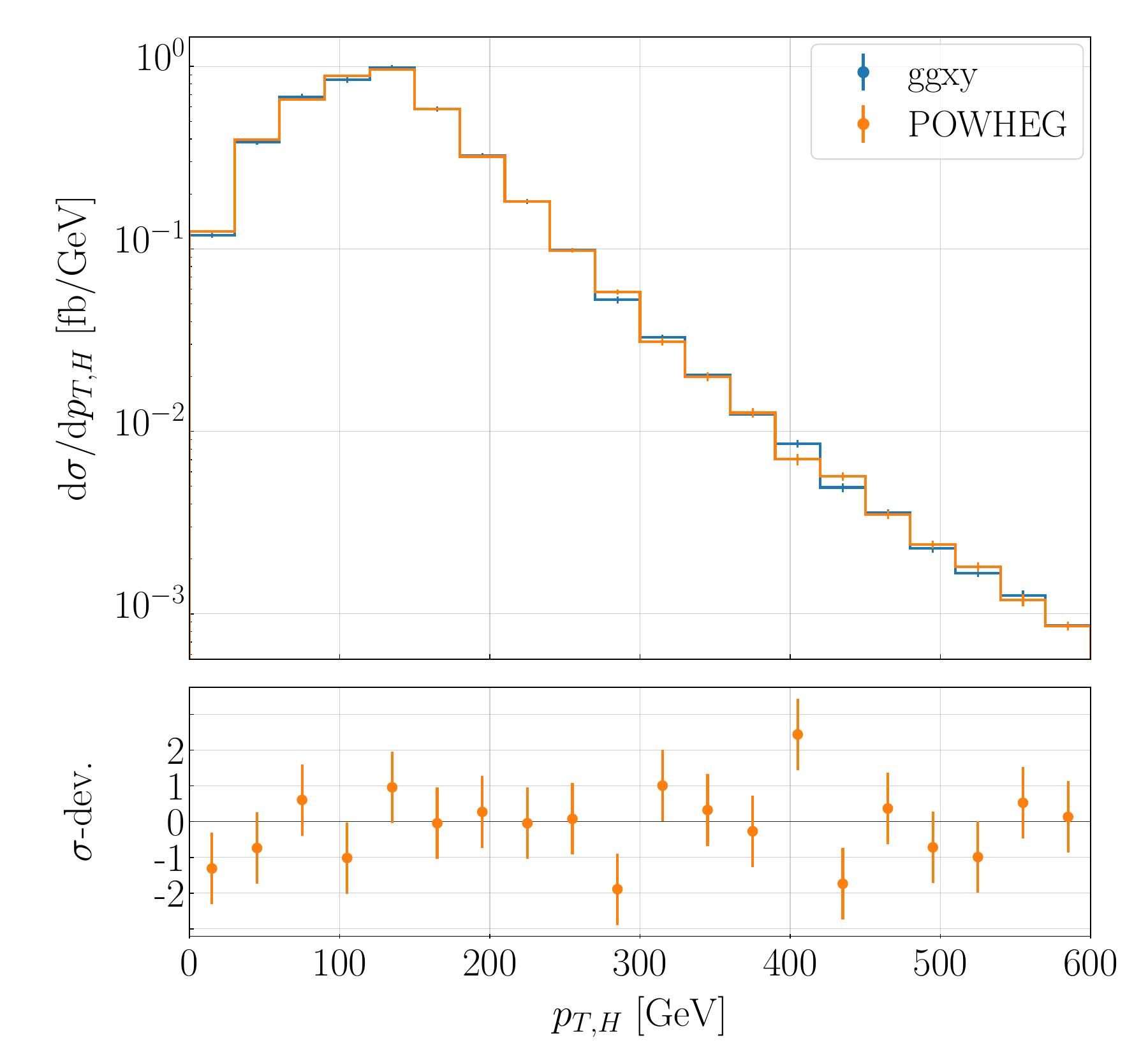}
\includegraphics[width=.32\textwidth]{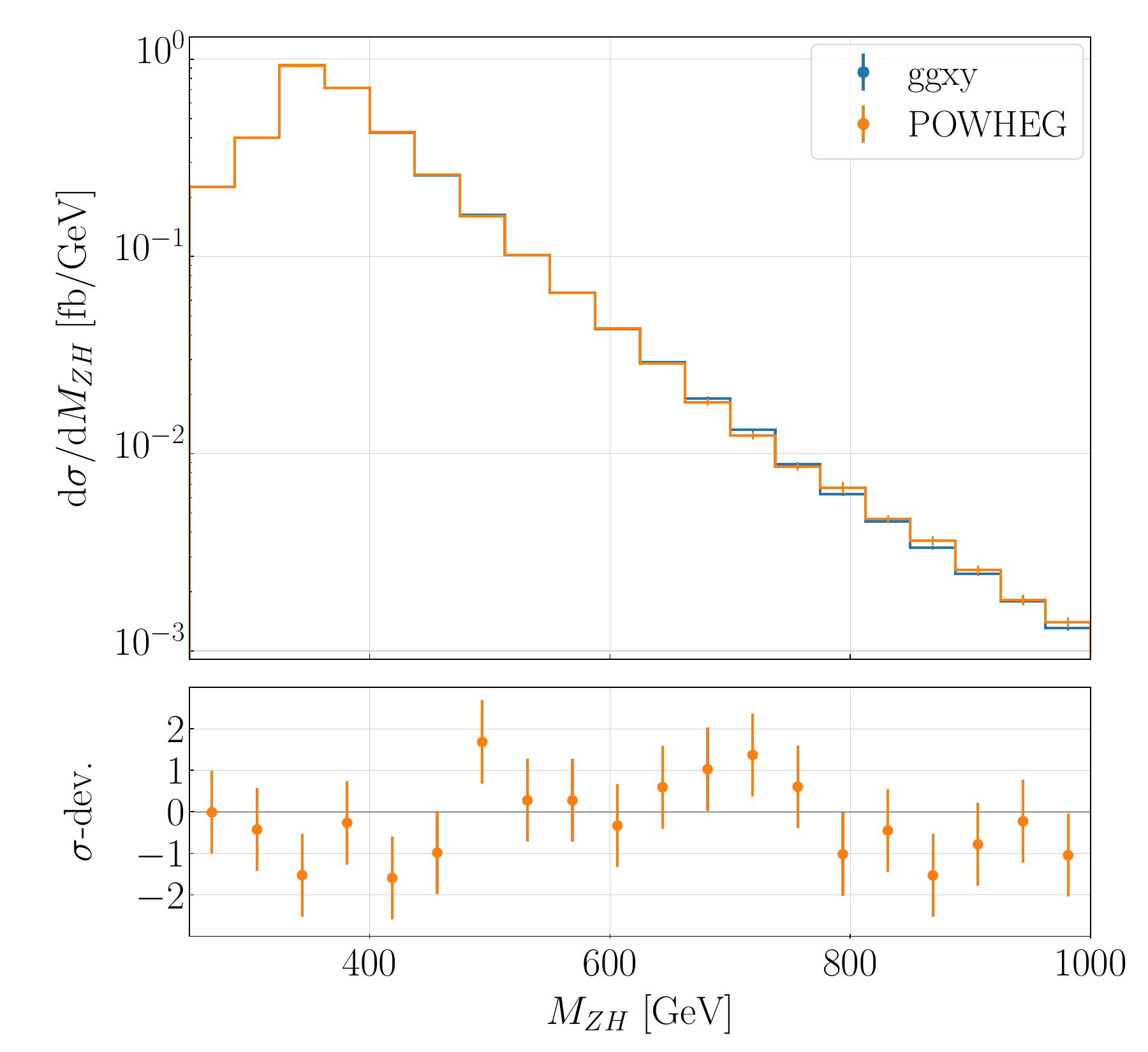}
\caption{\label{fig::compare_ggxy_pwhg} Differential distributions for the observables $p_{T,Z}$, $p_{T,H}$ and $M_{ZH}$ at NLO QCD for the $gg\to ZH$ process at $\sqrt{s}=13.6~{\rm TeV}$ computed with \texttt{ggxy} and the \texttt{POWHEG} interface.
In the lower panels we show the difference of the results normalized to the combined statistical uncertainties.}
\end{figure}
In addition, we have also cross-checked the two implementations at the differential level. As possible examples, we show in Fig.~\ref{fig::compare_ggxy_pwhg} the transverse momentum of the $Z$ and Higgs bosons as well as the invariant mass of the $ZH$ system. Overall the calculations agree very well.

\begin{figure}[t]
\centering
\includegraphics[width=.32\textwidth]{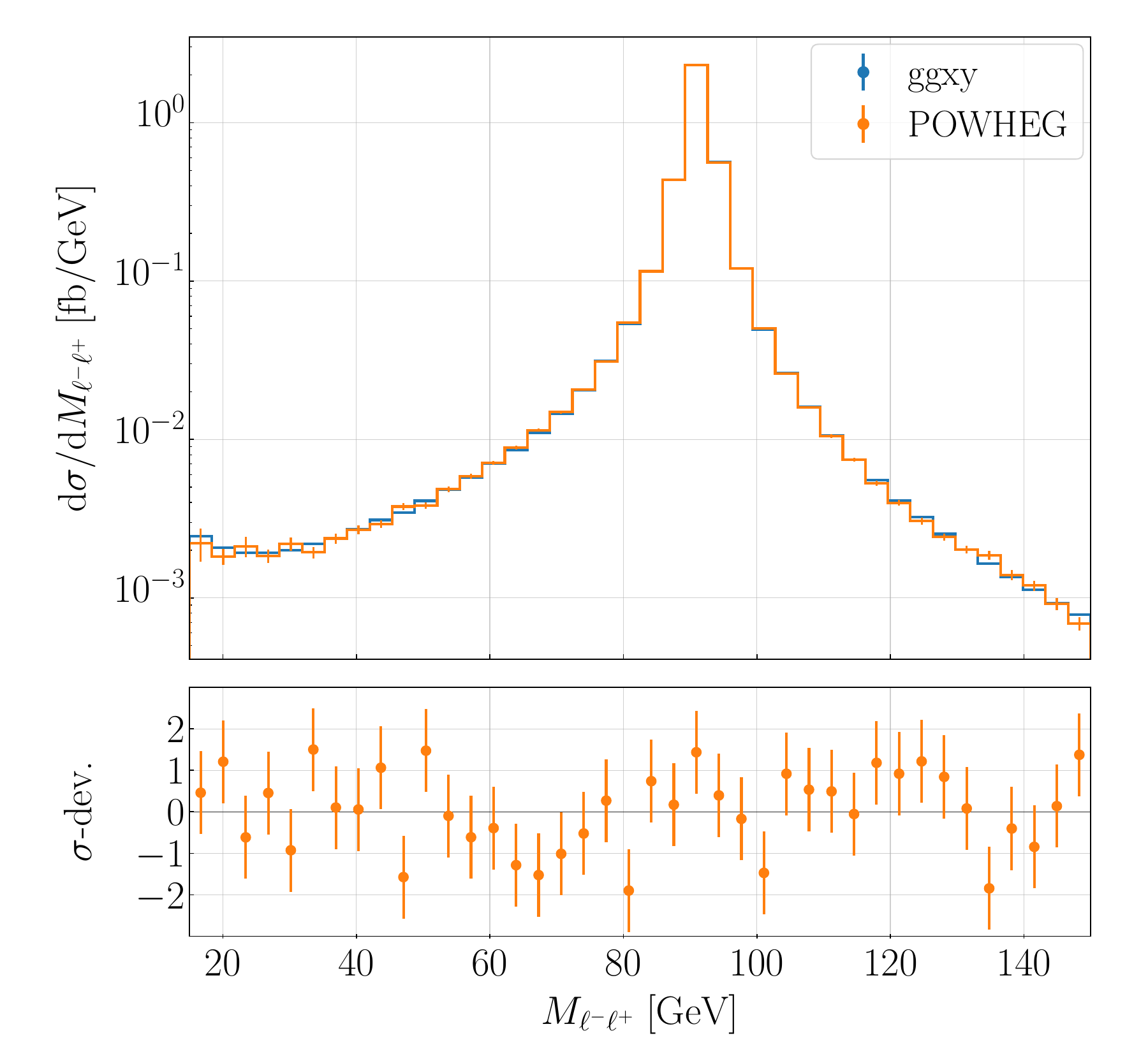}
\includegraphics[width=.32\textwidth]{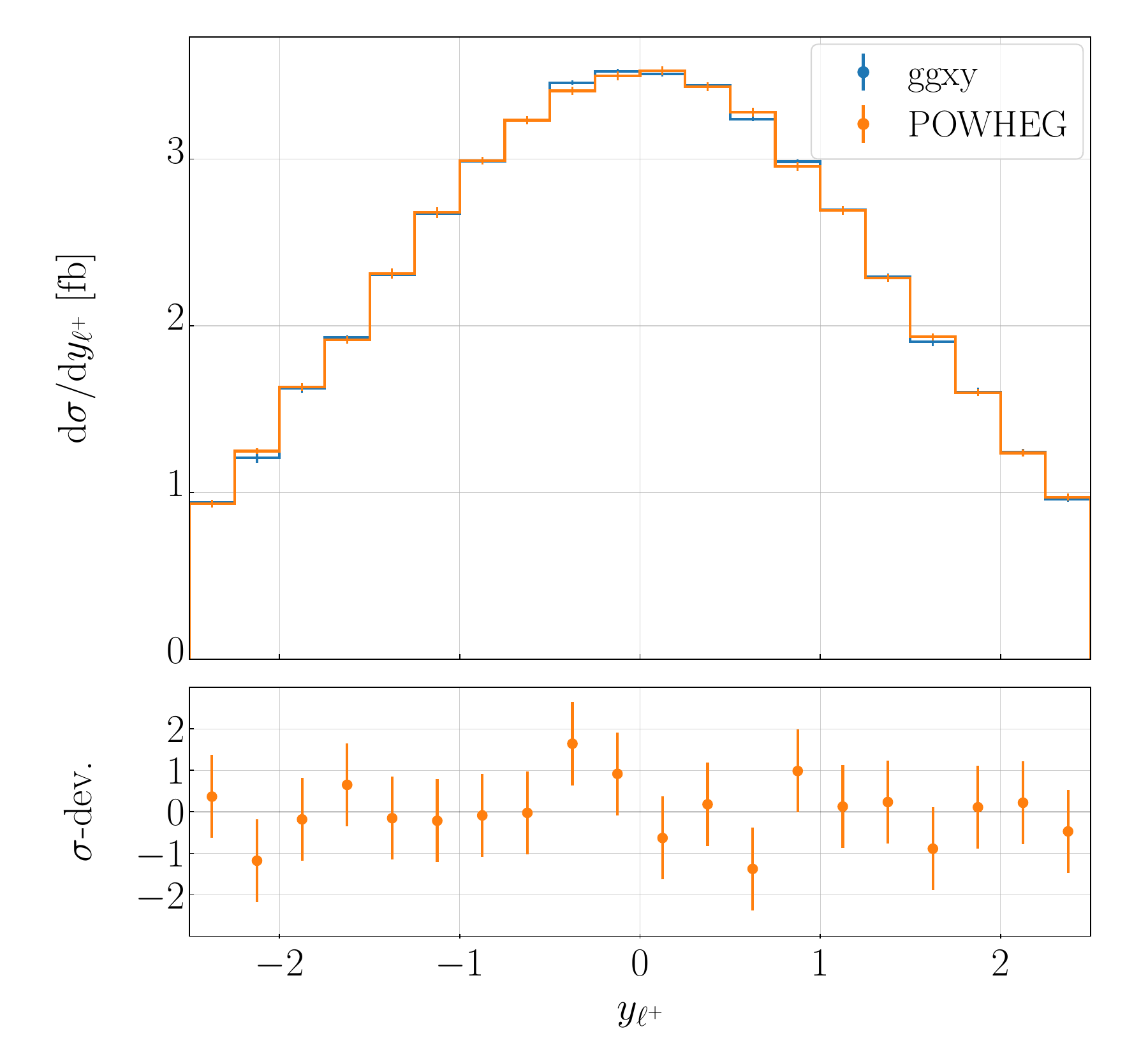}
\includegraphics[width=.32\textwidth]{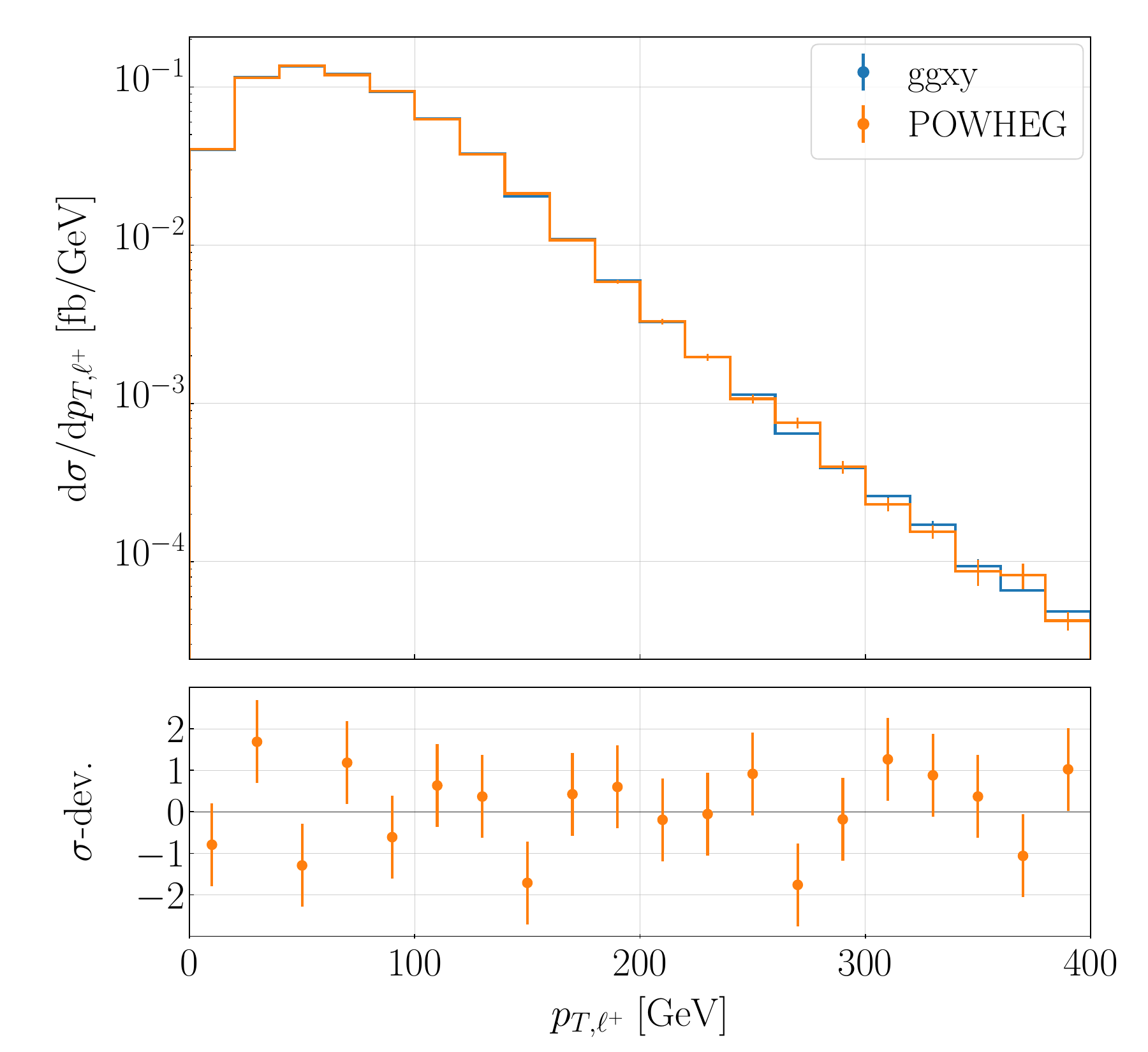}
\caption{\label{fig::compare_ggxy_pwhg-dec} Differential distributions in the observables $M_{\ell^-\ell^+}$, $y_{\ell^+}$ and $p_{T,\ell^+}$ at NLO QCD for the $gg\to \ell^-\ell^+ H$ process at $\sqrt{s}=13.6~{\rm TeV}$ computed with \texttt{ggxy} and the \texttt{POWHEG} interface.
In the lower panels we show the difference of the results normalized to the combined statistical uncertainties.}
\end{figure}

We have performed similar cross-checks for the $gg\to\ell^-\ell^+H$ process. Also in this case we reproduce the results at the level of the integrated cross section from the computation with \texttt{ggxy} from Tab.~\ref{tab::sig3}:
\begin{align*}
    \sigma^{\rm NLO}_{gg\to\ell^-\ell^+H}(\texttt{ggxy}) &= 13.02(1)~{\rm fb},\\
    \sigma^{\rm NLO}_{gg\to\ell^-\ell^+H}(\texttt{POWHEG}) &= 13.01(1)~{\rm fb}.
\end{align*}
Similarly good agreement is also found at the differential level. In Fig.~\ref{fig::compare_ggxy_pwhg-dec} we present, as an example, the differential cross sections w.r.t.~the invariant mass of the lepton pair, $M_{\ell^-\ell^+}$, the rapidity of the positively charged lepton $y_{\ell^+}$ and its transverse momentum $p_{T,\ell^-}$. In all cases we find good agreement between both calculations.

\subsection{\label{sub::num}Numerical results for parton showers with \texttt{Pythia}}

In this section we discuss parton shower effects and the modeling of the $Z$-boson decays at the differential level. For clarity, we consider the process the $gg\to e^-e^+H$. We use the parton shower of \texttt{Pythia} 8.2~\cite{Sjostrand:2007gs}; an example program (\verb|main-PYTHIA8.f|) can be found in the corresponding \texttt{POWHEG} implementation as well as an example run in the directory \verb|test-pythia8|. We use as default the same value for $\verb|hdamp|=250$ in \texttt{POWHEG} as for the calculation of $gg\to HH$ presented in Ref.~\cite{Heinrich:2017kxx}. This parameter can be used to divide the real corrections into two separate contributions $R=R_s+R_f$ with
\begin{align}
    R_s=F_{\rm damp} R, \qquad\qquad R_f=(1-F_{\rm damp}) R,
\end{align}
with
\begin{align}
    F_{\rm damp} = \frac{h_{\rm damp}^2}{h^2_{\rm damp}+k_T^2},
\end{align}
where $k_T^2$ is the transverse momentum of the radiation. The singular contribution $R_s$ is used in the matching procedure, while the finite contribution $R_f$ is treated as in a regular fixed-order calculation. The \texttt{POWHEG} default value corresponds to $\verb|hdamp|=\infty$, i.e., $F_{\rm damp}=1$. Changing this value corresponds to tuning the real cross section such that it follows the fixed-order results at large transverse momenta more closely, which becomes especially relevant in the case of large  
corrections from real radiation. Differences between the two choices will be discussed later in this section.

\begin{figure}[t]
\centering
\includegraphics[width=.49\textwidth]{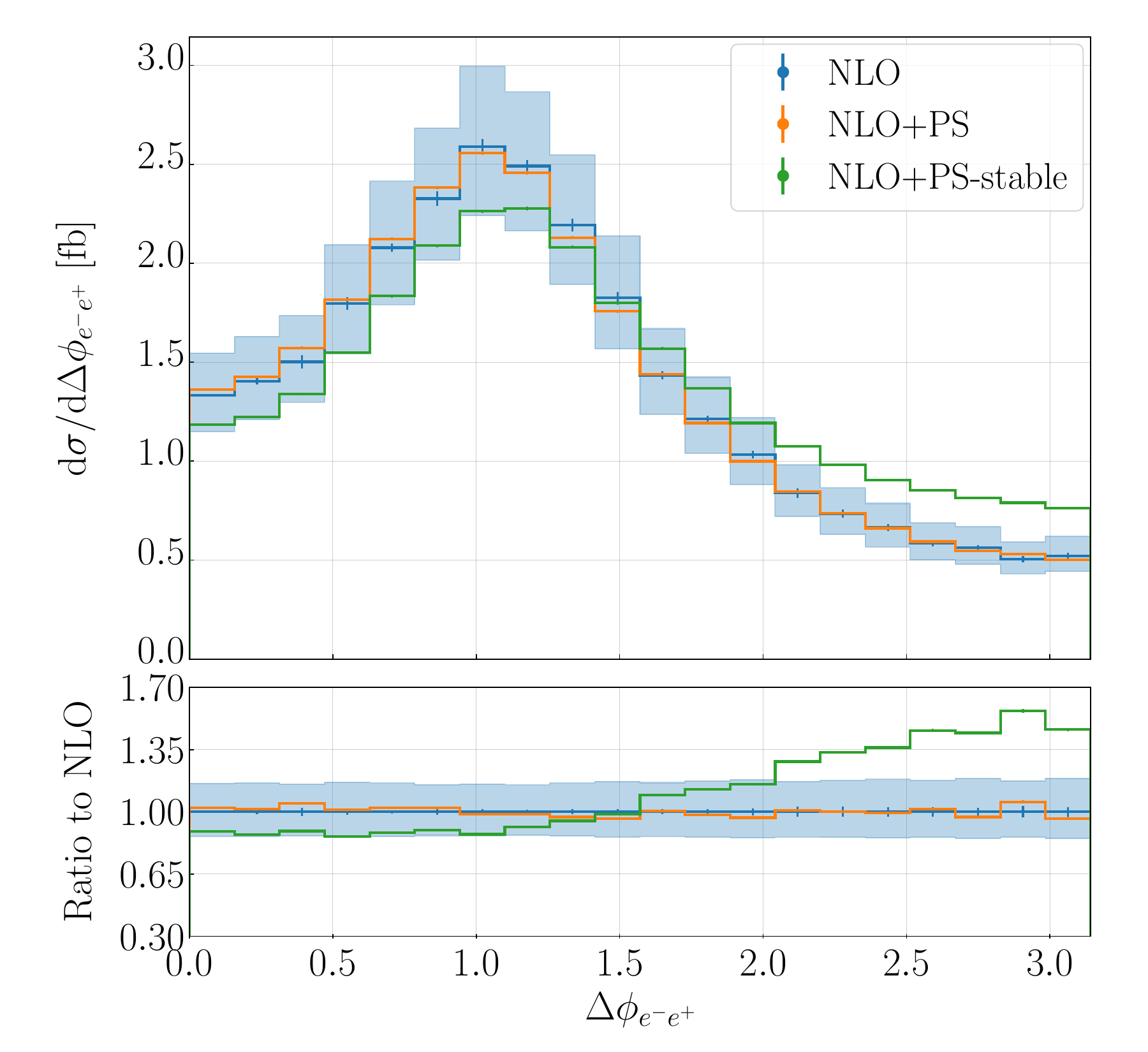}
\includegraphics[width=.49\textwidth]{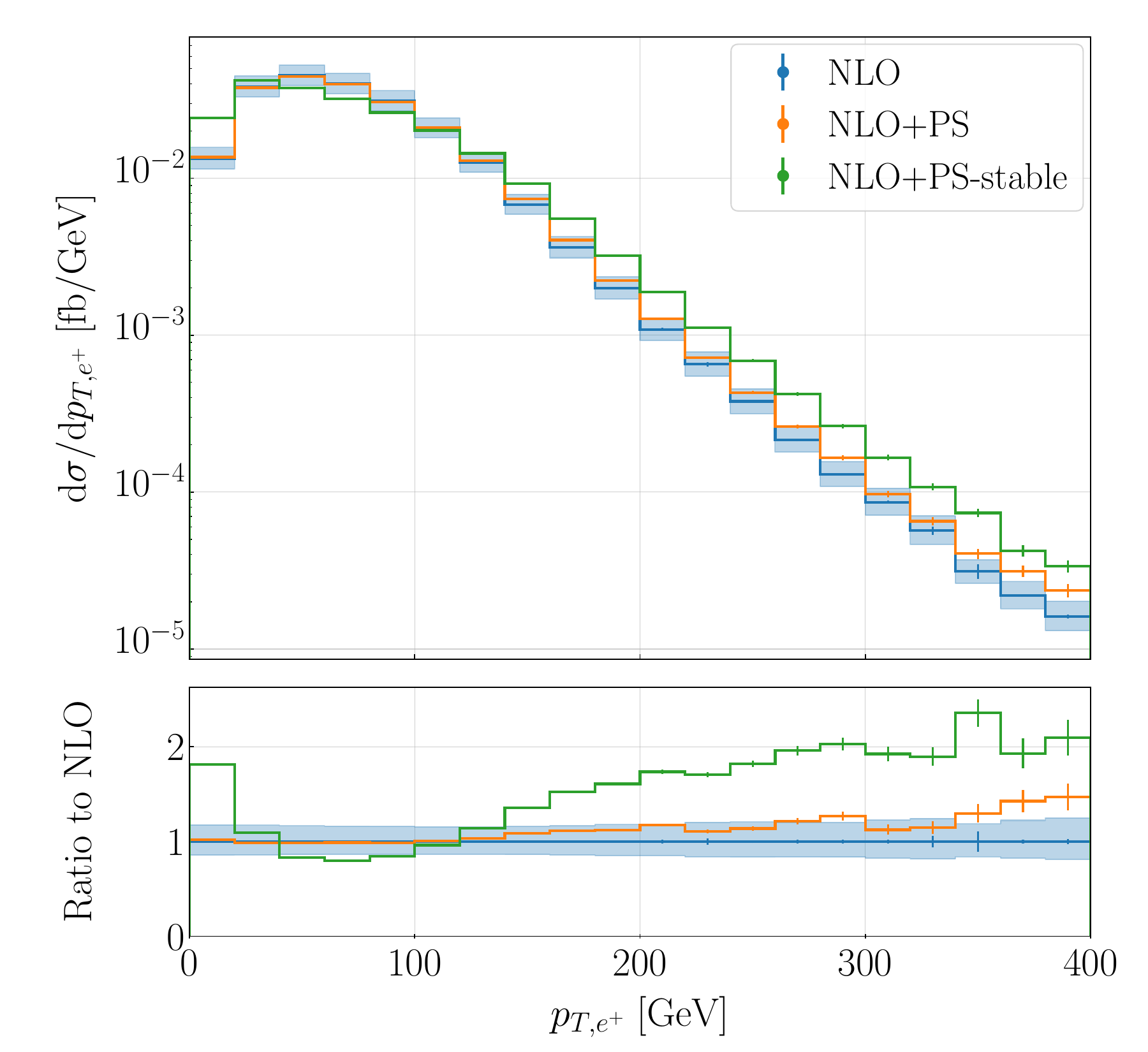}
\includegraphics[width=.49\textwidth]{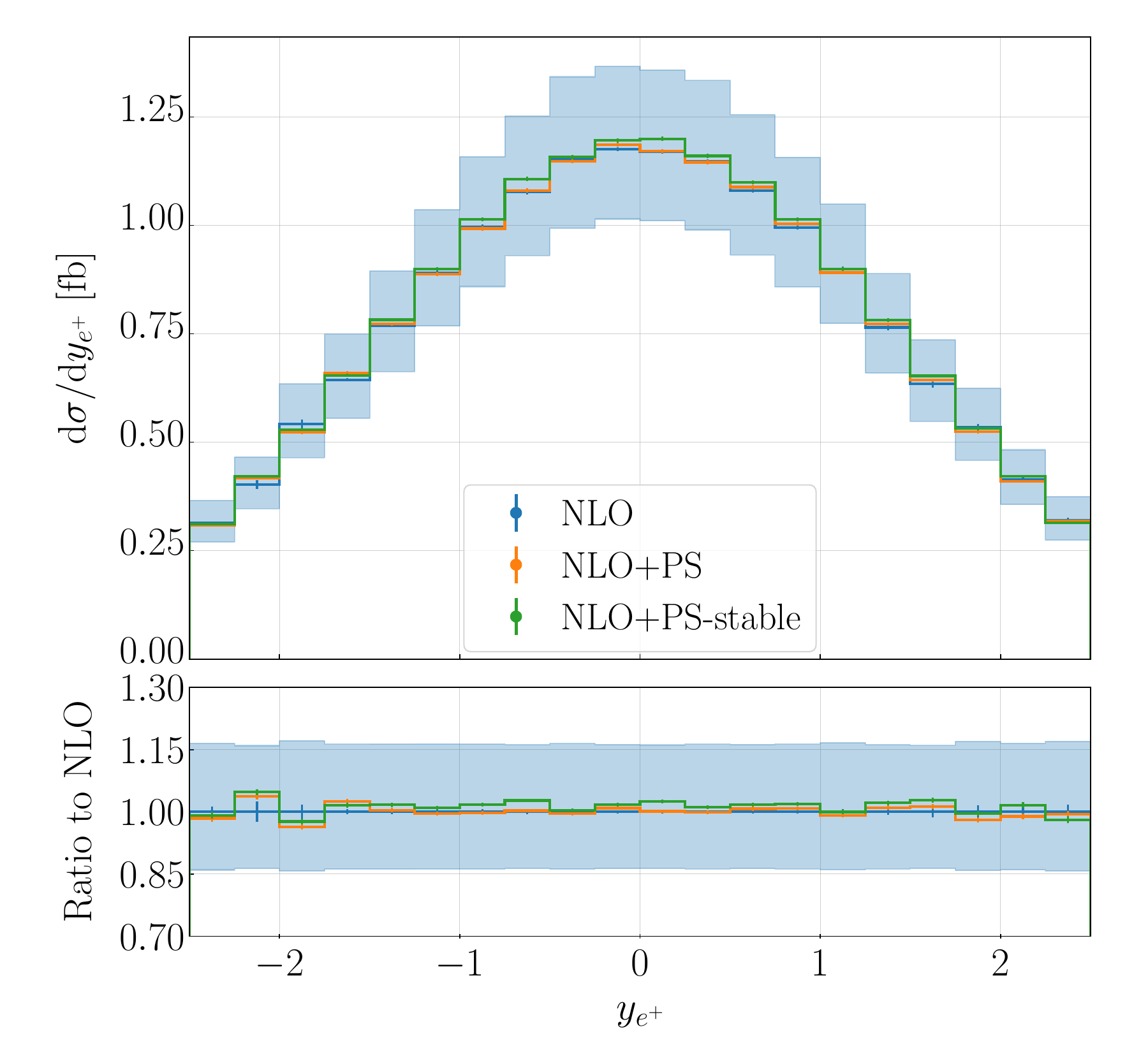}
\includegraphics[width=.49\textwidth]{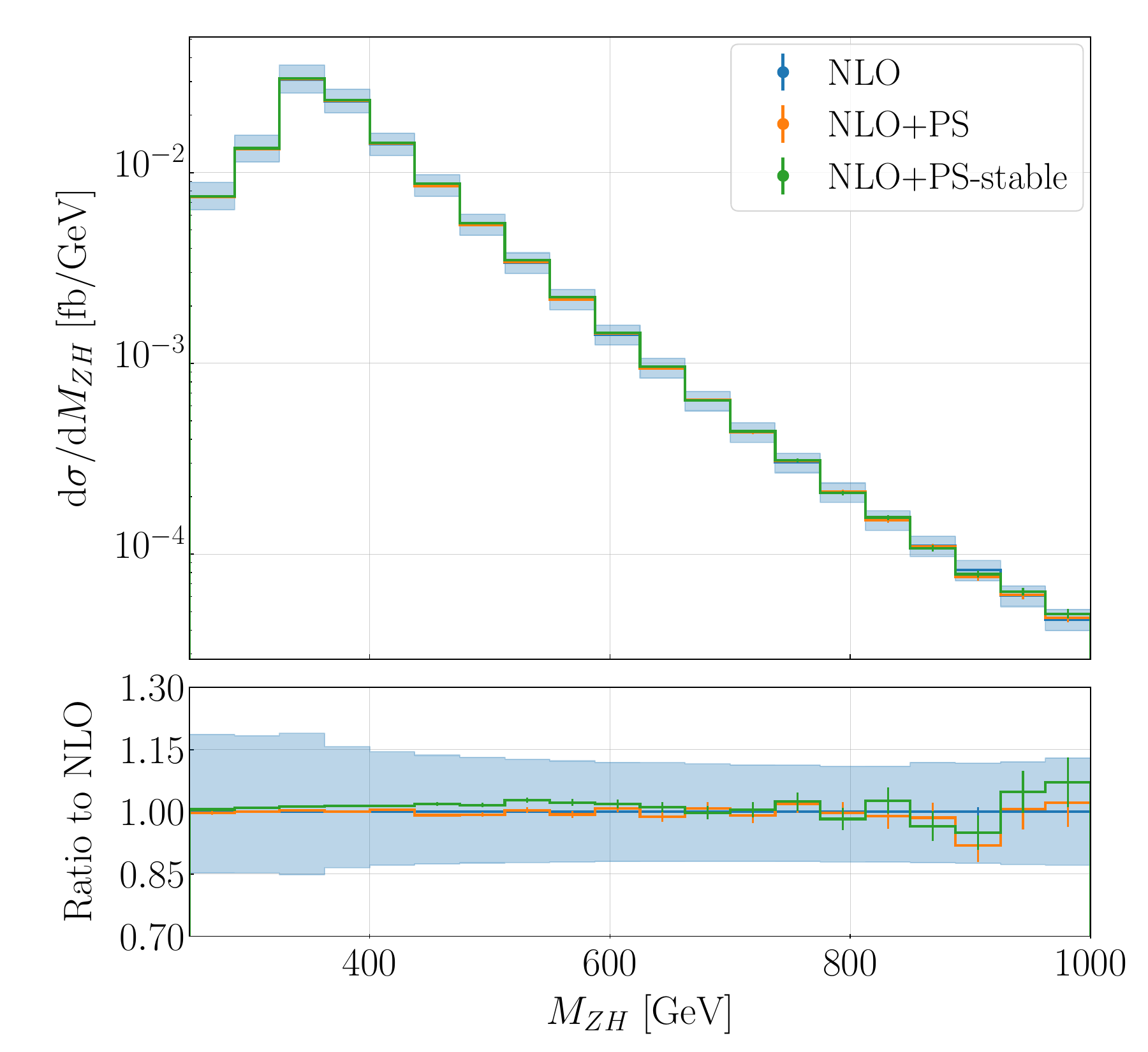}
\caption{\label{fig::compare_PS} Differential distributions for the observables $\Delta\phi_{e^-e^+}$, $p_{T,e^+}$, $y_{e^+}$ and $M_{ZH}$ for the $gg\to e^-e^+H$ process at NLO QCD. Results are shown at fixed-order (NLO), including parton shower effects with (NLO+PS) and without (NLO+PS-stable) $Z$ boson decays at the matrix element level.}
\end{figure}
In Fig.~\ref{fig::compare_PS} we compare the fixed-order predictions of the $gg\to e^-e^+H$ process at NLO QCD computed with \texttt{ggxy} (NLO) with two different predictions including parton shower effects. The two predictions matched to the parton shower of \texttt{Pythia} only differ in the modeling of the $Z$ boson, where the predictions labeled with ``NLO+PS'' contain, as the fixed-order results, the $Z$ boson decays already at the matrix element retaining all off-shell effects and spin correlations. The second prediction containing parton shower effects, ``NLO+PS-stable'', is generated with our second implementation for a stable $Z$ boson in \texttt{POWHEG}, where the $Z$ boson decays are simulated with \texttt{Pythia}. 

We find that, in general, the naive approach without spin correlations fails to precisely describe angular distributions of the two leptons such as the azimuthal angle between the electron-positron pair, $\Delta \phi_{e^-e^+}$, or the transverse momentum distribution of the positron $p_{T,e^+}$. The differences range from $40\%-100\%$ and are significantly larger than the scale uncertainties of the fixed-order calculation. On the other hand, the numerical results for the rapidity of the positron $y_{e^+}$ are essentially the same for all three approaches. This is the only leptonic observable for which we found that the inclusion of spin correlations is irrelevant. Also for observables that are built from the reconstructed $Z$ boson, the inclusion of spin correlations does not lead to any improvement as can be seen, e.g., from the invariant mass of the $ZH$ system $M_{ZH}$. Out of these four observables we find only for $p_{T,e^+}$ sizable parton shower effects, which increase towards the tails to more than $20\%$ and become similar in size to the corresponding scale uncertainties of the fixed-order calculation.

\begin{figure}[t]
\centering
\includegraphics[width=.49\textwidth]{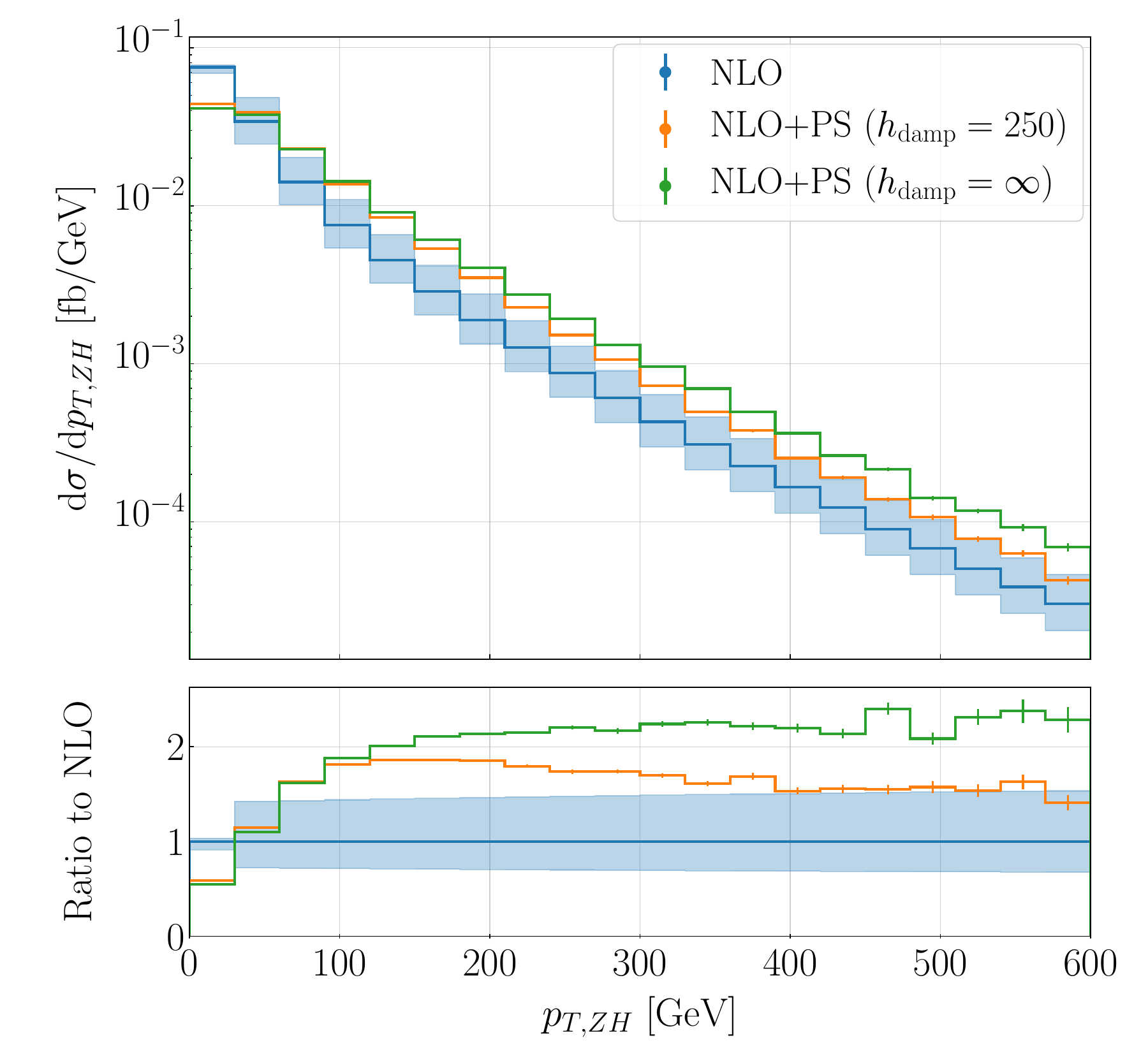}
\includegraphics[width=.49\textwidth]{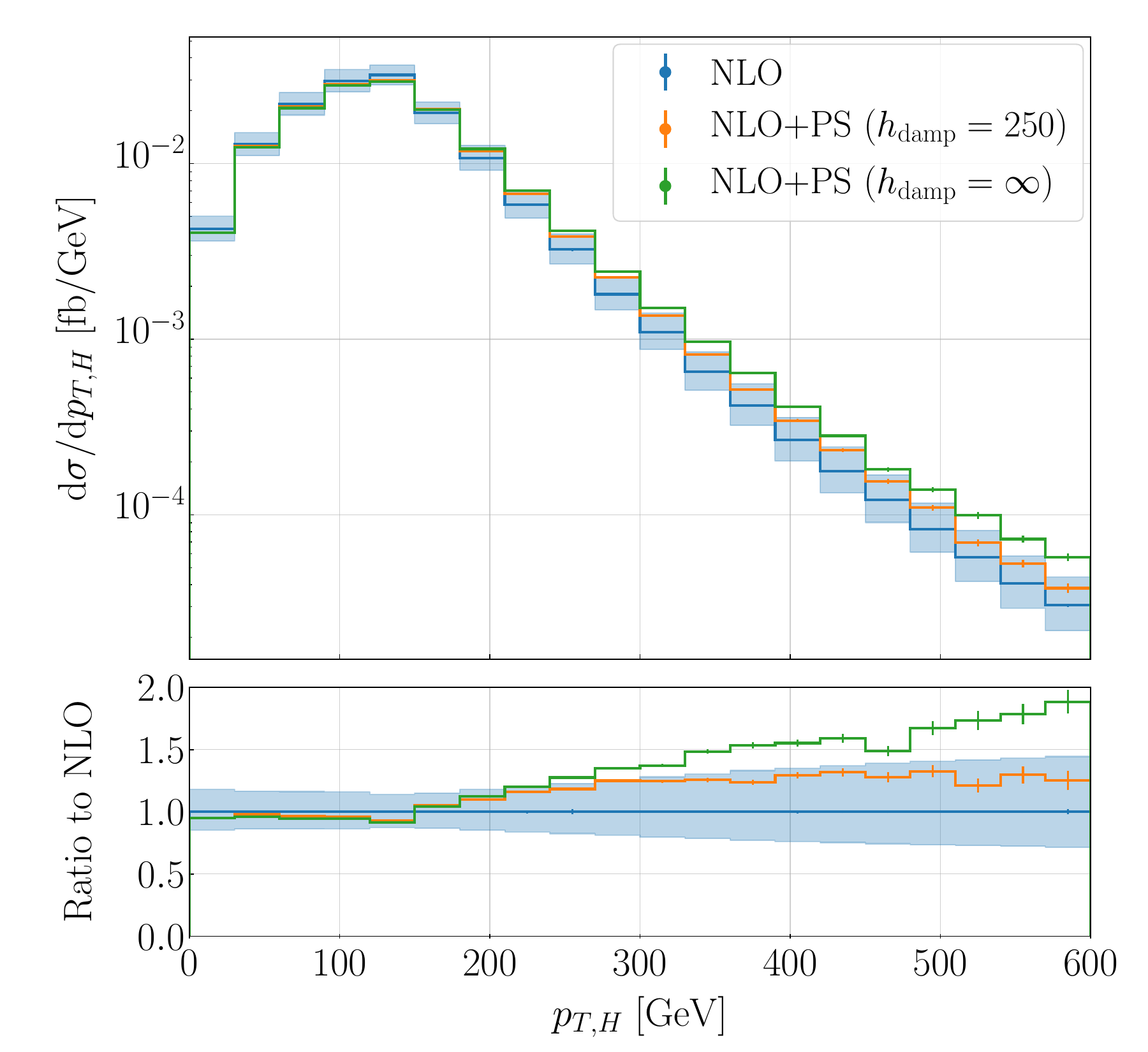}
\caption{\label{fig::compare_PS_hdamp}
Differential distributions for the observables $p_{T,ZH}$ and $p_{T,H}$ for the $gg\to e^-e^+H$ process at NLO QCD. Results are shown at fixed-order and including parton shower effects with $h_{\rm damp}=250$ and $h_{\rm damp}=\infty$.}
\end{figure}
In Fig.~\ref{fig::compare_PS_hdamp} we show the differential distributions of $p_{T,ZH}$ and $p_{T,H}$ for the $gg\to e^-e^+H$ at fixed-order as well as two results matched to parton showers, where we use for the \texttt{POWHEG} input parameter $h_{\rm damp}=250$ and $h_{\rm damp}=\infty$. These two observables have rather large parton shower effects compared to the rest that we have considered. For $p_{T,ZH}$ it is fairly clear that the inclusion of parton shower effects lead to significant shape changes, as at fixed-order (NLO) we have $p_{T,ZH}=p_{T,j}$, where $j$ is the additional radiation in the real corrections. We observe that the additional radiation from the parton shower leads to a softer distribution at small $p_{T,ZH}$ and to a harder  spectrum in the tail, with differences up to $100\%$ for $h_{\rm damp}=\infty$ with respect to the fixed-order prediction. 

For $p_{T,H}$ we find that the inclusion of parton shower effects leads to an increase of more than $80\%$ in the tail. One reason for such large effects is the size of the NLO corrections in this region, which are as large as $700\%$ (not shown in the panel), due to the inclusion of diagrams with $Z$ radiation off external quark lines in real corrections. For both distributions it is possible to choose a different value from $h_{\rm damp}=\infty$ such that the tails of the parton shower predictions are closer to the ones of the fixed-order results. With $h_{\rm damp}=250$ the size of the parton shower effects for both observables decrease by half to about $40\%-50\%$.

\section{\label{sec::concl}Conclusions}

In this paper we describe the implementation of NLO QCD corrections to
$gg\to ZH$ in the \texttt{C++} library \texttt{ggxy}.  It is based on
deep expansions in the large-$m_t$ limit, around the forward limit and
for high energies. \texttt{ggxy} provides a framework which allows for a
flexible choice of all input parameters. For the renormalization scheme of
the top quark mass one can either choose the on-shell or the
$\overline{\rm MS}$ scheme with arbitrary renormalization scale
$\mu_t$. Furthermore, it is possible to consider either stable $Z$
bosons or their leptonic decays into charged leptons of neutrinos
at the amplitude level.

We provide an interface to \texttt{POWHEG} which we use for
the matching to parton showers. The latter are realized with the
help of \texttt{Pythia}. This allows us to compare the effect of
$Z$ boson decays at the level of the amplitude to 
decays generated during the showering process.

The latest version of \texttt{ggxy} can be downloaded from
\url{https://gitlab.com/ggxy/ggxy-release}
and the \texttt{POWHEG} implementation from
\url{https://gitlab.com/POWHEG-BOX/V2/User-Processes/ggxy_ggZH}.

\section*{Acknowledgements}

This work was supported by the Deutsche Forschungsgemeinschaft (DFG, German
Research Foundation) under grant 396021762 --- TRR 257 ``Particle Physics
Phenomenology after the Higgs Discovery''.
The work of J.D.~was supported by STFC Consolidated Grant ST/X000699/1. 
K.S.~is supported by the European Union under the Marie Sk{\l}odowska-Curie Actions (MSCA) Grant 101204018.
We thank Benjamin Campillo
for providing results for the real corrections for individual phase-space points and for providing the data points for the observables
shown in Fig.~\ref{fig::comparison_ggxy_ZH}.

\bibliographystyle{jhep}

\bibliography{inspire.bib}

@article{ATLAS:2024yzu,
    author = "Aad, Georges and others",
    collaboration = "ATLAS",
    title = "{Measurements of WH and ZH production with Higgs boson decays into bottom quarks and direct constraints on the charm Yukawa coupling in 13 TeV pp collisions with the ATLAS detector}",
    eprint = "2410.19611",
    archivePrefix = "arXiv",
    primaryClass = "hep-ex",
    reportNumber = "CERN-EP-2024-237",
    doi = "10.1007/JHEP04(2025)075",
    journal = "JHEP",
    volume = "04",
    pages = "075",
    year = "2025"
}

@article{CMS:2023vzh,
    author = "Tumasyan, Armen and others",
    collaboration = "CMS",
    title = "{Measurement of simplified template cross sections of the Higgs boson produced in association with W or Z bosons in the H{\textrightarrow}bb{\textasciimacron} decay channel in proton-proton collisions at s=13{\,}{\,}TeV}",
    eprint = "2312.07562",
    archivePrefix = "arXiv",
    primaryClass = "hep-ex",
    reportNumber = "CMS-HIG-20-001, CERN-EP-2023-270",
    doi = "10.1103/PhysRevD.109.092011",
    journal = "Phys. Rev. D",
    volume = "109",
    number = "9",
    pages = "092011",
    year = "2024"
}

@article{Kumar:2014uwa,
    author = "Kumar, M. C. and Mandal, M. K. and Ravindran, V.",
    title = "{Associated production of Higgs boson with vector boson at threshold N$^{3}$LO in QCD}",
    eprint = "1412.3357",
    archivePrefix = "arXiv",
    primaryClass = "hep-ph",
    reportNumber = "HRI-RECAPP-2014-027",
    doi = "10.1007/JHEP03(2015)037",
    journal = "JHEP",
    volume = "03",
    pages = "037",
    year = "2015"
}

@article{Das:2025wbj,
    author = "Das, Goutam and Dey, Chinmoy and Kumar, M. C. and Samanta, Kajal",
    title = "{Soft gluon resummation for gluon fusion ZH production}",
    eprint = "2501.10330",
    archivePrefix = "arXiv",
    primaryClass = "hep-ph",
    reportNumber = "TTK-24-58, P3H-24-100, IPPP/24/81",
    doi = "10.1103/mxxw-bvm3",
    journal = "Phys. Rev. D",
    volume = "113",
    number = "1",
    pages = "014024",
    year = "2026"
}

@article{Harlander:2014wda,
    author = "Harlander, Robert V. and Kulesza, Anna and Theeuwes, Vincent and Zirke, Tom",
    title = "{Soft gluon resummation for gluon-induced Higgs Strahlung}",
    eprint = "1410.0217",
    archivePrefix = "arXiv",
    primaryClass = "hep-ph",
    reportNumber = "MS-TP-14-17, WUB-14-10, LPN14-112",
    doi = "10.1007/JHEP11(2014)082",
    journal = "JHEP",
    volume = "11",
    pages = "082",
    year = "2014"
}

@article{Davies:2020lpf,
    author = "Davies, Joshua and Mishima, Go and Steinhauser, Matthias and Wellmann, David",
    title = "{$gg\to ZZ$: analytic two-loop results for the low- and high-energy regions}",
    eprint = "2002.05558",
    archivePrefix = "arXiv",
    primaryClass = "hep-ph",
    reportNumber = "TTP20-006, P3H-20-007",
    doi = "10.1007/JHEP04(2020)024",
    journal = "JHEP",
    volume = "04",
    pages = "024",
    year = "2020"
}

@article{vanHameren:2010cp,
    author = "van Hameren, A.",
    title = "{OneLOop: For the evaluation of one-loop scalar functions}",
    eprint = "1007.4716",
    archivePrefix = "arXiv",
    primaryClass = "hep-ph",
    reportNumber = "IFJPAN-IV-2010-7",
    doi = "10.1016/j.cpc.2011.06.011",
    journal = "Comput. Phys. Commun.",
    volume = "182",
    pages = "2427--2438",
    year = "2011"
}

@article{Berends:1987me,
    author = "Berends, Frits A. and Giele, W. T.",
    title = "{Recursive Calculations for Processes with n Gluons}",
    reportNumber = "Print-88-0100 (LEIDEN)",
    doi = "10.1016/0550-3213(88)90442-7",
    journal = "Nucl. Phys. B",
    volume = "306",
    pages = "759--808",
    year = "1988"
}

@article{Degrassi:2022mro,
    author = {Degrassi, Giuseppe and Gr{\"o}ber, Ramona and Vitti, Marco and Zhao, Xiaoran},
    title = "{On the NLO QCD corrections to gluon-initiated ZH production}",
    eprint = "2205.02769",
    archivePrefix = "arXiv",
    primaryClass = "hep-ph",
    reportNumber = "CERN-TH-2022-079",
    doi = "10.1007/JHEP08(2022)009",
    journal = "JHEP",
    volume = "08",
    pages = "009",
    year = "2022"
}

@article{Actis:2016mpe,
    author = "Actis, Stefano and Denner, Ansgar and Hofer, Lars and Lang, Jean-Nicolas and Scharf, Andreas and Uccirati, Sandro",
    title = "{RECOLA: REcursive Computation of One-Loop Amplitudes}",
    eprint = "1605.01090",
    archivePrefix = "arXiv",
    primaryClass = "hep-ph",
    reportNumber = "ICCUB-16-020",
    doi = "10.1016/j.cpc.2017.01.004",
    journal = "Comput. Phys. Commun.",
    volume = "214",
    pages = "140--173",
    year = "2017"
}

@article{Denner:2016kdg,
    author = "Denner, Ansgar and Dittmaier, Stefan and Hofer, Lars",
    title = "{Collier: a fortran-based Complex One-Loop LIbrary in Extended Regularizations}",
    eprint = "1604.06792",
    archivePrefix = "arXiv",
    primaryClass = "hep-ph",
    reportNumber = "FR-PHENO-2016-003, ICCUB-16-016",
    doi = "10.1016/j.cpc.2016.10.013",
    journal = "Comput. Phys. Commun.",
    volume = "212",
    pages = "220--238",
    year = "2017"
}

@article{Buckley:2014ana,
    author = {Buckley, Andy and Ferrando, James and Lloyd, Stephen and Nordstr{\"o}m, Karl and Page, Ben and R{\"u}fenacht, Martin and Sch{\"o}nherr, Marek and Watt, Graeme},
    title = "{LHAPDF6: parton density access in the LHC precision era}",
    eprint = "1412.7420",
    archivePrefix = "arXiv",
    primaryClass = "hep-ph",
    reportNumber = "GLAS-PPE-2014-05, MCNET-14-29, IPPP-14-111, DCPT-14-222",
    doi = "10.1140/epjc/s10052-015-3318-8",
    journal = "Eur. Phys. J. C",
    volume = "75",
    pages = "132",
    year = "2015"
}

@article{PDF4LHCWorkingGroup:2022cjn,
    author = "Ball, Richard D. and others",
    collaboration = "PDF4LHC Working Group",
    title = "{The PDF4LHC21 combination of global PDF fits for the LHC Run III}",
    eprint = "2203.05506",
    archivePrefix = "arXiv",
    primaryClass = "hep-ph",
    reportNumber = "Edinburgh 2021/31, FERMILAB-PUB-22-121-QIS-SCD-T, MSUHEP-22-010, SMU-HEP-22-01, Nikhef 2021-033",
    doi = "10.1088/1361-6471/ac7216",
    journal = "J. Phys. G",
    volume = "49",
    number = "8",
    pages = "080501",
    year = "2022"
}

@article{Ossola:2006us,
    author = "Ossola, Giovanni and Papadopoulos, Costas G. and Pittau, Roberto",
    title = "{Reducing full one-loop amplitudes to scalar integrals at the integrand level}",
    eprint = "hep-ph/0609007",
    archivePrefix = "arXiv",
    doi = "10.1016/j.nuclphysb.2006.11.012",
    journal = "Nucl. Phys. B",
    volume = "763",
    pages = "147--169",
    year = "2007"
}

@article{Bellm:2025pcw,
    author = "Bellm, J. and others",
    title = "{The Physics of Herwig 7}",
    eprint = "2512.16645",
    archivePrefix = "arXiv",
    primaryClass = "hep-ph",
    reportNumber = "CERN-TH-2025-252, IPPP-25-57, HERWIG-2025-01, KA-TP-36-2025, MCNET-25-31",
    month = "12",
    year = "2025"
}

@article{Sjostrand:2007gs,
    author = "Sjostrand, Torbjorn and Mrenna, Stephen and Skands, Peter Z.",
    title = "{A Brief Introduction to PYTHIA 8.1}",
    eprint = "0710.3820",
    archivePrefix = "arXiv",
    primaryClass = "hep-ph",
    reportNumber = "CERN-LCGAPP-2007-04, LU-TP-07-28, FERMILAB-PUB-07-512-CD-T",
    doi = "10.1016/j.cpc.2008.01.036",
    journal = "Comput. Phys. Commun.",
    volume = "178",
    pages = "852--867",
    year = "2008"
}

@article{Alioli:2021wpn,
    author = {Alioli, Simone and Ferrario Ravasio, Silvia and Lindert, Jonas M. and R{\"o}ntsch, Raoul},
    title = "{Four-lepton production in gluon fusion at NLO matched to parton showers}",
    eprint = "2102.07783",
    archivePrefix = "arXiv",
    primaryClass = "hep-ph",
    reportNumber = "IPPP/20/79, CERN-TH-2021-022",
    doi = "10.1140/epjc/s10052-021-09470-5",
    journal = "Eur. Phys. J. C",
    volume = "81",
    number = "8",
    pages = "687",
    year = "2021"
}

@article{Alioli:2010xd,
    author = "Alioli, Simone and Nason, Paolo and Oleari, Carlo and Re, Emanuele",
    title = "{A general framework for implementing NLO calculations in shower Monte Carlo programs: the POWHEG BOX}",
    eprint = "1002.2581",
    archivePrefix = "arXiv",
    primaryClass = "hep-ph",
    reportNumber = "DESY-10-018, SFB-CPP-10-22, IPPP-10-11, DCPT-10-22",
    doi = "10.1007/JHEP06(2010)043",
    journal = "JHEP",
    volume = "06",
    pages = "043",
    year = "2010"
}

@article{Ossola:2007ax,
    author = "Ossola, Giovanni and Papadopoulos, Costas G. and Pittau, Roberto",
    title = "{CutTools: A Program implementing the OPP reduction method to compute one-loop amplitudes}",
    eprint = "0711.3596",
    archivePrefix = "arXiv",
    primaryClass = "hep-ph",
    doi = "10.1088/1126-6708/2008/03/042",
    journal = "JHEP",
    volume = "03",
    pages = "042",
    year = "2008"
}

@article{Actis:2012qn,
    author = "Actis, S. and Denner, A. and Hofer, L. and Scharf, A. and Uccirati, S.",
    title = "{Recursive generation of one-loop amplitudes in the Standard Model}",
    eprint = "1211.6316",
    archivePrefix = "arXiv",
    primaryClass = "hep-ph",
    reportNumber = "PSI-PR-12-09",
    doi = "10.1007/JHEP04(2013)037",
    journal = "JHEP",
    volume = "04",
    pages = "037",
    year = "2013"
}

@article{Catani:1996vz,
    author = "Catani, S. and Seymour, M. H.",
    title = "{A General algorithm for calculating jet cross-sections in NLO QCD}",
    eprint = "hep-ph/9605323",
    archivePrefix = "arXiv",
    reportNumber = "CERN-TH-96-029, CERN-TH-96-29",
    doi = "10.1016/S0550-3213(96)00589-5",
    journal = "Nucl. Phys. B",
    volume = "485",
    pages = "291--419",
    year = "1997",
    note = "[Erratum: Nucl.Phys.B 510, 503--504 (1998)]"
}

@article{Davies:2025qjr,
    author = {Davies, Joshua and Sch{\"o}nwald, Kay and Steinhauser, Matthias and Stremmer, Daniel},
    title = "{ggxy: A flexible library to compute gluon-induced cross sections}",
    eprint = "2506.04323",
    archivePrefix = "arXiv",
    primaryClass = "hep-ph",
    reportNumber = "LTH 1406, P3H-25-035, TTP25-017, ZU-TH40/25",
    doi = "10.1016/j.cpc.2025.109933",
    journal = "Comput. Phys. Commun.",
    volume = "320",
    pages = "109933",
    year = "2026"
}

@inproceedings{CampilloAveleira:2025rbh,
    author = "Campillo Aveleira, Benjamin and others",
    title = "{$gg \to ZH$ : updated predictions at NLO QCD}",
    eprint = "2508.09905",
    archivePrefix = "arXiv",
    primaryClass = "hep-ph",
    reportNumber = "LHCHWG-2025-007, COMETA-2025-31, TTP25-026, KA-TP-22-2025, P3H-25-057",
    month = "8",
    year = "2025"
}

@article{Ferrera:2014lca,
    author = "Ferrera, Giancarlo and Grazzini, Massimiliano and Tramontano, Francesco",
    title = "{Associated ZH production at hadron colliders: the fully differential NNLO QCD calculation}",
    eprint = "1407.4747",
    archivePrefix = "arXiv",
    primaryClass = "hep-ph",
    reportNumber = "IFUM-1031-FT, ZU-TH-23-14",
    doi = "10.1016/j.physletb.2014.11.040",
    journal = "Phys. Lett. B",
    volume = "740",
    pages = "51--55",
    year = "2015"
}

@article{Chen:2022rua,
    author = "Chen, Long and Davies, Joshua and Heinrich, Gudrun and Jones, Stephen P. and Kerner, Matthias and Mishima, Go and Schlenk, Johannes and Steinhauser, Matthias",
    title = "{ZH production in gluon fusion at NLO in QCD}",
    eprint = "2204.05225",
    archivePrefix = "arXiv",
    primaryClass = "hep-ph",
    reportNumber = "IPPP/22/19,P3H-22-038,KA-TP-08-2022,TTP22-024,TU-1147,PSI-PR-22-08",
    doi = "10.1007/JHEP08(2022)056",
    journal = "JHEP",
    volume = "08",
    pages = "056",
    year = "2022"
}

@article{Altenkamp:2012sx,
    author = "Altenkamp, Lukas and Dittmaier, Stefan and Harlander, Robert V. and Rzehak, Heidi and Zirke, Tom J. E.",
    title = "{Gluon-induced Higgs-strahlung at next-to-leading order QCD}",
    eprint = "1211.5015",
    archivePrefix = "arXiv",
    primaryClass = "hep-ph",
    reportNumber = "CERN-PH-TH-2012-312, FR-PHENO-2012-023, WUB-12-21",
    doi = "10.1007/JHEP02(2013)078",
    journal = "JHEP",
    volume = "02",
    pages = "078",
    year = "2013"
}

@article{Borowka:2016ypz,
    author = "Borowka, S. and Greiner, N. and Heinrich, G. and Jones, S. P. and Kerner, M. and Schlenk, J. and Zirke, T.",
    title = "{Full top quark mass dependence in Higgs boson pair production at NLO}",
    eprint = "1608.04798",
    archivePrefix = "arXiv",
    primaryClass = "hep-ph",
    reportNumber = "MPP-2016-261, ZU-TH-31-16",
    doi = "10.1007/JHEP10(2016)107",
    journal = "JHEP",
    volume = "10",
    pages = "107",
    year = "2016"
}

@article{Harlander:2013mla,
    author = "Harlander, Robert V. and Liebler, Stefan and Zirke, Tom",
    title = "{Higgs Strahlung at the Large Hadron Collider in the 2-Higgs-Doublet Model}",
    eprint = "1307.8122",
    archivePrefix = "arXiv",
    primaryClass = "hep-ph",
    reportNumber = "WUB-13-12",
    doi = "10.1007/JHEP02(2014)023",
    journal = "JHEP",
    volume = "02",
    pages = "023",
    year = "2014"
}

@article{Alasfar:2021ppe,
    author = {Alasfar, Lina and Degrassi, Giuseppe and Giardino, Pier Paolo and Gr{\"o}ber, Ramona and Vitti, Marco},
    title = "{Virtual corrections to $gg\to ZH$ via a transverse momentum expansion}",
    eprint = "2103.06225",
    archivePrefix = "arXiv",
    primaryClass = "hep-ph",
    doi = "10.1007/JHEP05(2021)168",
    journal = "JHEP",
    volume = "05",
    pages = "168",
    year = "2021"
}

@article{Denner:2014cla,
    author = {Denner, Ansgar and Dittmaier, Stefan and Kallweit, Stefan and M{\"u}ck, Alexander},
    title = "{HAWK  2.0: A Monte Carlo program for Higgs production in vector-boson fusion and Higgs strahlung at hadron colliders}",
    eprint = "1412.5390",
    archivePrefix = "arXiv",
    primaryClass = "hep-ph",
    reportNumber = "FR-PHENO-2014-013, MITP-14-101, TTK-14-36",
    doi = "10.1016/j.cpc.2015.04.021",
    journal = "Comput. Phys. Commun.",
    volume = "195",
    pages = "161--171",
    year = "2015"
}

@article{Harlander:2018yio,
    author = "Harlander, Robert V. and Klappert, Jonas and Liebler, Stefan and Simon, Lukas",
    title = "{vh@nnlo-v2: New physics in Higgs Strahlung}",
    eprint = "1802.04817",
    archivePrefix = "arXiv",
    primaryClass = "hep-ph",
    reportNumber = "KA-TP-01-2018, TTK-17-47",
    doi = "10.1007/JHEP05(2018)089",
    journal = "JHEP",
    volume = "05",
    pages = "089",
    year = "2018"
}

@article{Baglio:2022wzu,
    author = "Baglio, Julien and Duhr, Claude and Mistlberger, Bernhard and Szafron, Robert",
    title = "{Inclusive production cross sections at N$^{3}$LO}",
    eprint = "2209.06138",
    archivePrefix = "arXiv",
    primaryClass = "hep-ph",
    reportNumber = "CERN-TH-2022-109, SLAC-PUB-17699, BONN-TH-2022-22",
    doi = "10.1007/JHEP12(2022)066",
    journal = "JHEP",
    volume = "12",
    pages = "066",
    year = "2022"
}

@article{Dicus:1988yh,
    author = "Dicus, Duane A. and Kao, Chung",
    title = "{Higgs Boson - $Z^0$ Production From Gluon Fusion}",
    reportNumber = "DOE-ER40200-131",
    doi = "10.1103/PhysRevD.38.1008",
    journal = "Phys. Rev. D",
    volume = "38",
    pages = "1008",
    year = "1988",
    note = "[Erratum: Phys.Rev.D 42, 2412 (1990)]"
}

@article{Ciccolini:2003jy,
    author = "Ciccolini, M. L. and Dittmaier, S. and Kramer, M.",
    title = "{Electroweak radiative corrections to associated WH and ZH production at hadron colliders}",
    eprint = "hep-ph/0306234",
    archivePrefix = "arXiv",
    reportNumber = "EDINBURGH-2003-05, MPI-PHT-2003-24",
    doi = "10.1103/PhysRevD.68.073003",
    journal = "Phys. Rev. D",
    volume = "68",
    pages = "073003",
    year = "2003"
}

@article{Brein:2011vx,
    author = "Brein, Oliver and Harlander, Robert and Wiesemann, Marius and Zirke, Tom",
    title = "{Top-Quark Mediated Effects in Hadronic Higgs-Strahlung}",
    eprint = "1111.0761",
    archivePrefix = "arXiv",
    primaryClass = "hep-ph",
    reportNumber = "CERN-PH-TH-2011-268, FR-PHENO-2011-016, WUB-11-15",
    doi = "10.1140/epjc/s10052-012-1868-6",
    journal = "Eur. Phys. J. C",
    volume = "72",
    pages = "1868",
    year = "2012"
}

@article{Kniehl:1990iva,
    author = "Kniehl, Bernd A.",
    title = "{Associated Production of Higgs and Z Bosons From Gluon Fusion in Hadron Collisions}",
    reportNumber = "MAD-PH-558",
    doi = "10.1103/PhysRevD.42.2253",
    journal = "Phys. Rev. D",
    volume = "42",
    pages = "2253--2258",
    year = "1990"
}

@article{Chen:2020gae,
    author = "Chen, Long and Heinrich, Gudrun and Jones, Stephen P. and Kerner, Matthias and Klappert, Jonas and Schlenk, Johannes",
    title = "{$ZH$ production in gluon fusion: two-loop amplitudes with full top quark mass dependence}",
    eprint = "2011.12325",
    archivePrefix = "arXiv",
    primaryClass = "hep-ph",
    reportNumber = "ZU-TH 45/20, CERN-TH-2020-199, IPPP/20/57, P3H-20-076,
  KA-TP-21-2020, P3H-20-076, KA-TP-21-2020, TTK-20-42, PSI-PR-20-21, PSI-PR-20-21 P3H-20-076{\textbackslash}{\textbackslash} KA-TP-21-2020{\textbackslash}{\textbackslash} TTK-20-42{\textbackslash}{\textbackslash} PSI-PR-20-21",
    doi = "10.1007/JHEP03(2021)125",
    journal = "JHEP",
    volume = "03",
    pages = "125",
    year = "2021"
}

@article{Brein:2012ne,
    author = "Brein, Oliver and Harlander, Robert V. and Zirke, Tom J. E.",
    title = "{vh@nnlo - Higgs Strahlung at hadron colliders}",
    eprint = "1210.5347",
    archivePrefix = "arXiv",
    primaryClass = "hep-ph",
    doi = "10.1016/j.cpc.2012.11.002",
    journal = "Comput. Phys. Commun.",
    volume = "184",
    pages = "998--1003",
    year = "2013"
}

@article{Wang:2021rxu,
    author = "Wang, Guoxing and Xu, Xiaofeng and Xu, Yongqi and Yang, Li Lin",
    title = "{Next-to-leading order corrections for $gg\to ZH$ with top quark mass dependence}",
    eprint = "2107.08206",
    archivePrefix = "arXiv",
    primaryClass = "hep-ph",
    doi = "10.1016/j.physletb.2022.137087",
    journal = "Phys. Lett. B",
    volume = "829",
    pages = "137087",
    year = "2022"
}

@article{Harlander:2002wh,
    author = "Harlander, Robert V. and Kilgore, William B.",
    title = "{Next-to-next-to-leading order Higgs production at hadron colliders}",
    eprint = "hep-ph/0201206",
    archivePrefix = "arXiv",
    reportNumber = "BNL-HET-02-3, CERN-TH-2002-006",
    doi = "10.1103/PhysRevLett.88.201801",
    journal = "Phys. Rev. Lett.",
    volume = "88",
    pages = "201801",
    year = "2002"
}

@article{Hamberg:1990np,
    author = "Hamberg, R. and van Neerven, W. L. and Matsuura, T.",
    title = "{A complete calculation of the order $\alpha-s^{2}$ correction to the Drell-Yan $K$ factor}",
    reportNumber = "DESY-90-129",
    doi = "10.1016/0550-3213(91)90064-5",
    journal = "Nucl. Phys. B",
    volume = "359",
    pages = "343--405",
    year = "1991",
    note = "[Erratum: Nucl.Phys.B 644, 403--404 (2002)]"
}

@article{Han:1991ia,
    author = "Han, Tao and Willenbrock, S.",
    title = "{QCD correction to the p p ---{\ensuremath{>}} W H and Z H total cross-sections}",
    reportNumber = "FERMILAB-PUB-91-070-T, BNL-45990",
    doi = "10.1016/0370-2693(91)90572-8",
    journal = "Phys. Lett. B",
    volume = "273",
    pages = "167--172",
    year = "1991"
}

@article{Davies:2025out,
    author = {Davies, Joshua and Grau, Dominik and Sch{\"o}nwald, Kay and Steinhauser, Matthias and Stremmer, Daniel and Vitti, Marco},
    title = "{Two-loop QCD corrections to $ZH$ and off-shell $Z$ boson pair production in gluon fusion}",
    eprint = "2509.07072",
    archivePrefix = "arXiv",
    primaryClass = "hep-ph",
    reportNumber = "P3H-25-060, TTP25-029, ZU-TH 56/25",
    month = "9",
    year = "2025"
}

@article{Davies:2025otz,
    author = {Davies, Joshua and Grau, Dominik and Sch{\"o}nwald, Kay and Steinhauser, Matthias and Stremmer, Daniel},
    title = "{Three-loop corrections to $gg\to ZH$ in the large top quark mass limit}",
    eprint = "2512.00156",
    archivePrefix = "arXiv",
    primaryClass = "hep-ph",
    reportNumber = "CERN-TH-2025-245, P3H-25-101, TTP25-047",
    month = "11",
    year = "2025"
}

@article{Heinrich:2017kxx,
    author = "Heinrich, G. and Jones, S. P. and Kerner, M. and Luisoni, G. and Vryonidou, E.",
    title = "{NLO predictions for Higgs boson pair production with full top quark mass dependence matched to parton showers}",
    eprint = "1703.09252",
    archivePrefix = "arXiv",
    primaryClass = "hep-ph",
    reportNumber = "CERN-TH-2017-069, MPP-2017-41, NIKHEF-2017-020",
    doi = "10.1007/JHEP08(2017)088",
    journal = "JHEP",
    volume = "08",
    pages = "088",
    year = "2017"
}

@article{Davies:2020drs,
    author = "Davies, Joshua and Mishima, Go and Steinhauser, Matthias",
    title = "{Virtual corrections to $gg\to ZH$ in the high-energy and large-$m_t$ limits}",
    eprint = "2011.12314",
    archivePrefix = "arXiv",
    primaryClass = "hep-ph",
    reportNumber = "TTP20-041, P3H-20-074",
    doi = "10.1007/JHEP03(2021)034",
    journal = "JHEP",
    volume = "03",
    pages = "034",
    year = "2021"
}

@article{Brein:2003wg,
    author = "Brein, Oliver and Djouadi, Abdelhak and Harlander, Robert",
    title = "{NNLO QCD corrections to the Higgs-strahlung processes at hadron colliders}",
    eprint = "hep-ph/0307206",
    archivePrefix = "arXiv",
    reportNumber = "MPP-2003-35, CERN-TH-2003-161, PM-03-16",
    doi = "10.1016/j.physletb.2003.10.112",
    journal = "Phys. Lett. B",
    volume = "579",
    pages = "149--156",
    year = "2004"
}

@article{Hasselhuhn:2016rqt,
    author = "Hasselhuhn, Alexander and Luthe, Thomas and Steinhauser, Matthias",
    title = "{On top quark mass effects to $gg\to ZH$ at NLO}",
    eprint = "1611.05881",
    archivePrefix = "arXiv",
    primaryClass = "hep-ph",
    reportNumber = "TTP16-051",
    doi = "10.1007/JHEP01(2017)073",
    journal = "JHEP",
    volume = "01",
    pages = "073",
    year = "2017"
}

@article{Campbell:2016jau,
    author = "Campbell, John M. and Ellis, R. Keith and Williams, Ciaran",
    title = "{Associated production of a Higgs boson at NNLO}",
    eprint = "1601.00658",
    archivePrefix = "arXiv",
    primaryClass = "hep-ph",
    reportNumber = "IPPP-15-78, FERMILAB-PUB-16-001-T",
    doi = "10.1007/JHEP06(2016)179",
    journal = "JHEP",
    volume = "06",
    pages = "179",
    year = "2016"
}

@article{Denner:2011id,
    author = "Denner, Ansgar and Dittmaier, Stefan and Kallweit, Stefan and Muck, Alexander",
    title = "{Electroweak corrections to Higgs-strahlung off W/Z bosons at the Tevatron and the LHC with HAWK}",
    eprint = "1112.5142",
    archivePrefix = "arXiv",
    primaryClass = "hep-ph",
    reportNumber = "FR-PHENO-2011-025, PSI-PR-11-04, ZU-TH-29-11, TTK-11-61",
    doi = "10.1007/JHEP03(2012)075",
    journal = "JHEP",
    volume = "03",
    pages = "075",
    year = "2012"
}

\end{document}